\documentclass[pdflatex,sn-mathphys-num]{sn-jnl}
\usepackage{graphicx}%
\usepackage{multirow}%
\usepackage{amsmath,amssymb,amsfonts}%
\usepackage{amsthm}%
\usepackage{mathrsfs}%
\usepackage[title]{appendix}%
\usepackage{xcolor}%
\usepackage{textcomp}%
\usepackage{manyfoot}%
\usepackage{booktabs}%
\usepackage{threeparttable}%
\usepackage{algorithm}%
\usepackage{algorithmicx}%
\usepackage{algpseudocode}%
\usepackage{hyperref}
\usepackage{multirow}
\usepackage{chemformula}
\usepackage[version=3]{mhchem} 
\usepackage{caption}
\usepackage{subcaption}
\usepackage{makecell}
\usepackage{array}

\usepackage{siunitx}
\DeclareSIUnit{\angstrom}{\text{\AA}}
\DeclareSIUnit{\atom}{atom}

\begin{document}

\title[Article Title]{Reweighting free energy profiles between universal machine learning interatomic potentials for fast consensus building}

\author[1]{\fnm{Sauradeep} \sur{Majumdar}}
\author[1]{\fnm{Miguel} \sur{Steiner}} 
\author*[2]{\fnm{Johannes C. B.} \sur{Dietschreit}\email{johannes.dietschreit@univie.ac.at}} 
\author[1]{\fnm{Swagata} \sur{Roy}} 
\author[3]{\fnm{Daniel} \sur{Willimetz}}
\author[3]{\fnm{Lukaš} \sur{Grajciar}}
\author*[1]{\fnm{Rafael} \sur{G\'omez-Bombarelli}\email{rafagb@mit.edu}} 

\affil*[1]{\orgdiv{Department of Material Science and Engineering}, \orgname{Massachusetts Institute of Technology}, \orgaddress{\city{Cambridge}, \state{Massachusetts}, \postcode{02139}, \country{United States}}}
\affil[2]{\orgdiv{Institute of Theoretical Chemistry}, \orgname{University of Vienna}, \orgaddress{\street{Währinger Str. 17}, \postcode{1090}, \city{Vienna}, \country{Austria}}}
\affil[3]{\orgdiv{Department of Physical and Macromolecular Chemistry}, \orgname{Charles University}, \orgaddress{\street{Hlavova 8}, \city{Praha 2}, \postcode{12800}, \country{Czech Republic}}}


\abstract{From understanding complex chemical reaction mechanisms to determining transport kinetics of ions and gases in materials, free energy profiles serve as a fundamental bridge between microscopic atomic fluctuations and macroscopic thermodynamic observables. Estimating the free energy profile along a reaction coordinate, referred to as the potential of mean force (PMF), with density functional theory (DFT) accuracy is computationally expensive. The advent of universal machine learning interatomic potentials (MLIPs) has reduced this cost drastically while maintaining sufficient accuracy. However, the accuracy of an MLIP is strongly determined by its training data and hence can be uncertain for a given system. 
In this work, we present a systematic and scalable framework for reweighting PMFs, initially sampled with a single `source' MLIP, across a representative suite of target MLIPs. 
Because traditional direct exponential reweighting fails for large system sizes due to low phase-space overlap between potentials, we deploy robust analytical corrections. 
Applying this to a complex 601-atom system of \ce{Li+} transport in a nanoconfined electrolyte, we demonstrate that a mean energy-gap approximation effectively bypasses statistical collapse, producing a highly stable PMF that closely matches the target PMF. Using this approach, we recover high-fidelity target thermodynamics across multiple DFT reference levels (PBE+D3, PBE-sol, r\textsuperscript{2}SCAN, r\textsuperscript{2}SCAN-D4) at a fraction of the computational cost of full simulations. 
Furthermore, thermodynamic analysis reveals that the studied MLIPs partition into two distinct clusters driven by their training data. Our reweighting framework successfully recovers target thermodynamic properties--specifically, reaction free energy and activation free energy--even when the phase-space overlap between potentials is critically low. Ultimately, the demonstrated reweighting approach establishes a vital diagnostic protocol to achieve affordable cross-model consensus on materials chemistry properties without redundant, resource-intensive simulations.}

\keywords{statistical mechanics, machine learning interatomic potentials, enhanced sampling, nanoporous materials, ion transport}

\maketitle
\section{Introduction}\label{sec1}

Molecular dynamics (MD) simulations have become a cornerstone for gaining fundamental atomistic insights in chemical and material sciences\cite{frenkel2023understanding}.
However, many processes of interest, such as reaction pathways, binding events, and phase transitions, are rare events that cannot be observed on the timescales affordable with conventional MD\cite{henin2022enhanced}.
To overcome this limitation, enhanced sampling techniques such as umbrella sampling or metadynamics introduce a bias along a chosen reaction coordinate to drive the system across the relevant barriers\cite{tiwary2016review, yang2019enhanced, shen2023enhanced}.
The potential of mean force (PMF) along this coordinate is then recovered from the biased trajectories with techniques such as the Multistate Bennett Acceptance Ratio (MBAR)\cite{shirts2008statistically} or the Weighted Histogram Analysis Method (WHAM)\cite{kumar1992weighted}, providing the quantitative backbone necessary to understand the underlying chemistry\cite{chipot2007free, dietschreit2023entropy, millan2023}.
While these methods are robust, even generating the underlying biased MD trajectories is computationally expensive for large systems (hundreds or more of atoms) when density functional theory (DFT) levels of accuracy are required.

The recent advent of universal machine learning interatomic potentials (MLIPs)\cite{batatia2025foundation, wood2025family, rhodes2025orb, mazitov2025pet, yuan2026foundation} provides a powerful opportunity to tackle this challenge. 
We previously demonstrated the efficacy of MLIP-driven umbrella sampling in modeling Li- and Na-ion transport in nanoporous materials, achieving DFT-level accuracy at a fraction of the cost for length- and time-scales of up to 1000 atoms and 45~ns, respectively\cite{majumdar2026free}. 
However, the rapid proliferation of new MLIP architectures poses a logistical bottleneck: it is increasingly infeasible to `brute-force' extensive PMF calculations for every new and improved MLIP. 
Furthermore, different MLIPs often exhibit disparate computational profiles, with some models optimized for speed and lower memory footprints, while others prioritize maximum physical fidelity. 
It would thus be highly advantageous to perform extensive MD sampling with a computationally lean MLIP, and subsequently reweight those trajectories with free energy perturbation (FEP) \cite{zwanzig1954high} to obtain the PMF of a more accurate, yet more expensive, target potential. 
In its standard form, FEP relates the free energy difference $\Delta F_{\mathrm{A} \to \mathrm{B}}$ between a reference potential $U_A$ and a target potential $U_B$ via
\begin{equation}
    \Delta F_{\mathrm{A} \to \mathrm{B}} = -k_\mathrm{B} T\, \ln \left\langle \exp\!\left[-\beta\,(U_\mathrm{B} - U_\mathrm{A})\right] \right\rangle_\mathrm{A},
    \label{eq:fep}
\end{equation}
where $\beta = 1/(k_\mathrm{B} T)$ and the ensemble average is taken over configurations sampled with the reference potential A.
Assessing the consistency of PMFs across multiple MLIPs in this manner not only circumvents redundant sampling but also significantly enhances the fidelity and cross-model reliability of the results.

However, direct exponential reweighting between disparate energy models is fundamentally challenged by the strict requirement of phase-space overlap between the reference and target ensembles\cite{konig2014, hudson2019use, pohorille2010good}.
Because the variance of the exponential weight in eqn.~\eqref{eq:fep} grows linearly with the number of degrees of freedom\cite{wu2005phase}, exact direct reweighting is practically intractable for realistic system sizes (see the example for an $n$-dimensional harmonic oscillator model in SI Section~S7).
While traditional exponential averaging might succeed for small molecules in isolation, applying it to complex systems such as ion transport in solid-state electrolytes or gas diffusion in nanoporous materials inevitably pushes the FEP calculation into a regime where statistical collapse is expected. 
We treat this limitation as a motivation for the present work. 
By leveraging an information-theoretic diagnostic, the reweighting entropy\cite{Li2018} (see eqn.~\eqref{eq:reweighting_entropy}), we explicitly quantify this regime of phase-space collapse. 
Subsequently, we demonstrate that adopting an energy-only approximation (neglecting the highly noisy entropic contribution to the free energy correction) provides a robust, practical path forward. 
By bypassing the severe statistical noise of the exact exponential estimator, this approach successfully recovers the underlying free energy landscapes and enables a rigorous comparison of thermodynamic driving forces across a suite of universal MLIPs.
For a mathematical backdrop, see Eqs.~\eqref{eq:direct_reweighting}-\eqref{eq:energy_only} in the Methods~\ref{sec:methods} section.

We demonstrate this reweighting methodology by applying it to a system of interest for energy storage applications: \ce{Li+} transport in nanoconfinement\cite{cai2023unravelling,fong2025physical,majumdar2026free}. Specifically, we study \ce{Li+} diffusion through the nanopores of a chabazite (CHA) zeolite\cite{Prez-Botella2022, millan2023} in the presence of solvent (water in this case), with a total system size of 601 atoms.
We propose a computational workflow that uses a single `source' MLIP simulation (MACE-MP0)\cite{batatia2025foundation} to obtain PMFs for a representative suite of target MLIPs through reweighting, from UMA\cite{wood2025family}, MACE\cite{mace_github}, ORB\cite{rhodes2025orb}, and UPET\cite{mazitov2025pet} universal potentials.
To validate our approach, we carry out an umbrella sampling simulation from scratch for one of the most structurally divergent target MLIPs (MACE-MATPES) and confirm our reweighting recovers the target PMF. 
Ultimately, we highlight the scenarios under which cross-MLIP reweighting can provide fast, reliable estimates of thermodynamic properties of practical interest, and establish clear guidelines for when carrying out an entirely new simulation can be beneficial.


\section{Results}
\subsection{Reweighting strategy}

\begin{figure}
    \centering
    \includegraphics[width=\linewidth]{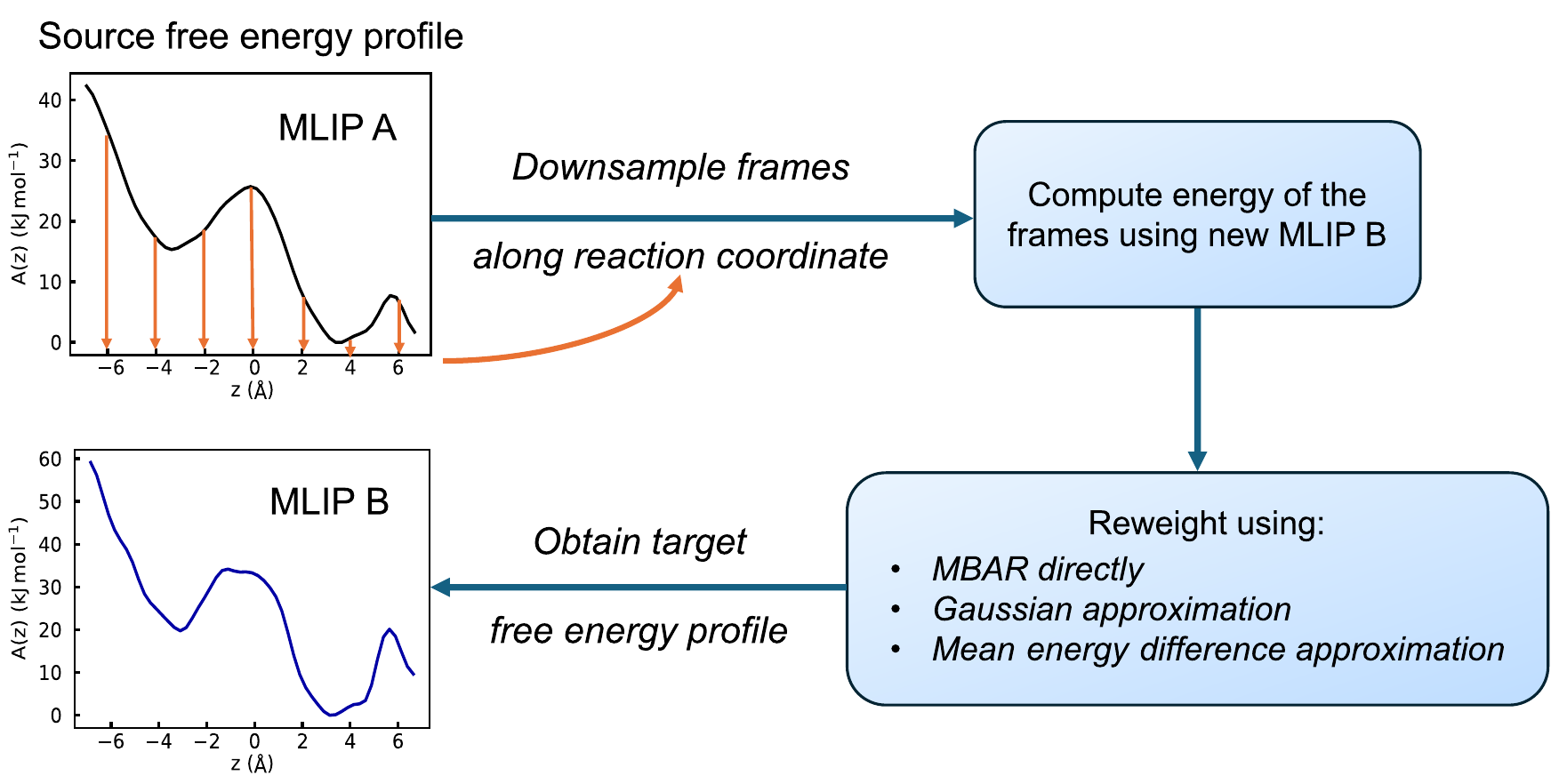}
    \caption{Illustration of the reweighting strategy used in this work.
    As the system, we use \ce{Li+} diffusion through an aluminum-doped zeolite (CHA), which has been studied by umbrella sampling in our previous work\cite{majumdar2026free}. 
    We take 10,000 frames per umbrella window from the source simulations (MLIP A) and evaluate their energies with the target potential (MLIP B). 
    Reweighting of these frames to obtain the target PMF was carried out with the estimator best suited based on the variance of the potential energy differences.}
    \label{fig:schematic}
\end{figure}

Figure~\ref{fig:schematic} illustrates the reweighting strategy adopted in this work. 
The starting PMF of the studied system, \ce{Li+} diffusing through the pore channel of a water-solvated chabazite (CHA) zeolite, has been obtained from our previous work\cite{majumdar2026free} (see Figure~S1 for an illustration of the studied system). 
The collective variable (CV) for the system is defined as the projection of the vector connecting the midpoint between the centers of mass of the two outer 8-membered rings (forming the pore channel) and the center of mass of \ce{Li+} onto the axis of diffusion. Thus, the zero point of the CV is at the center of the pore channel connecting the two 8-membered rings\cite{millan2023}.
We see a global minimum between 2.5~\AA\, and 5~\AA, which is due to the interaction between the mobile \ce{Li+} and the \ce{AlO4-} unit located in one of the outer 8-membered rings. 
MACE-MP0\cite{batatia2025foundation} (MLIP A in this work) was used to compute the source PMF.

10,000 (25 \%) frames were sampled from each of the 54 umbrella sampling windows. 
This ensures that the resultant 540,000 frames, each with 601 atoms, are distributed along the CV scale (i.e., along the pore channel) and are not localized at a particular point. 
Figures~S2b and~c show that the downsampled frames provide sufficient overlap between the umbrella windows, which is crucial for `stitching' the PMF together. 
Consequently, Figure~S3 shows that the 540,000 downsampled frames are well representative of the entire $\approx$ 2.1 million-frame MACE-MP0 trajectory. 
The energy of these downsampled frames was then calculated with the target MLIP~B, and the PMF corresponding to MLIP~B was obtained by reweighting. 
The downsampling of frames thus reduces the required number of evaluations of MLIP~B. 
Together with the possibility of running evaluations with multiple processes and leveraging batch evaluations of multiple frames at once on graphical processing units, this reduces the computational costs massively compared to running an entire biased MD simulation with MLIP~B.


\subsection{Assessing reweighting success and limitations}

\begin{table*}[ht]
\small\centering
\caption{Summary of the universal MLIPs evaluated in this work, including their training datasets, DFT level of theory, and MAEs for forces and energies evaluated against DFT (PBE+D3) calculations on 1100 Li-zeolite-\ce{H2O} configurations  (each with 601 atoms).}
\label{tab:mlips}
\begin{tabular}{@{}p{3.4cm}
                p{2.7cm}@{\hspace{1pt}}
                p{1.9cm}
                c
                c@{}}
\toprule
\textbf{MLIP}
    & \textbf{Training} \newline \textbf{Dataset}
    & \textbf{Functional}
    & \multicolumn{1}{p{1.9cm}}{\textbf{Force MAE} \newline \textbf{(\si{\kilo\joule\per\mole\per\angstrom})}}
    & \multicolumn{1}{p{2.3cm}@{}}{\textbf{Energy MAE}$^{b}$ \newline \textbf{(\si{\kilo\joule\per\mole\per\atom})}} \\
\midrule
MACE-MP0~\cite{batatia2025foundation}
    & MPTrj
    & PBE
    &  6.04 & 0.07 \\[4pt]
UMA~\cite{wood2025family}
    & OMAT-24
    & PBE
    &  1.81 & 0.02 \\[4pt]
PET-MAD-1.0.2~\cite{mazitov2025pet}
    & MAD
    & PBE-sol
    &  9.10 & 0.15 \\[4pt]
ORB-v3~\cite{rhodes2025orb}
    & OMAT-24
    & PBE
    &  2.72 & 0.04 \\[4pt]
MACE-MP0-D3~\cite{batatia2025foundation, takamoto2022towards}
    & MPTrj
    & PBE + D3$^{a}$
    &  6.02 & 0.14 \\[4pt]
MACE-MPA-0~\cite{mace_github}
    & MPTrj + sAlex
    & PBE
    &  4.61 & 0.09 \\[4pt]
MACE-MATPES~\cite{mace_github}
    & MATPES
    & r\textsuperscript{2}SCAN
    & 15.47 & 0.08 \\[4pt]
MACE-fine-tuned
    & zeolite~\cite{Erlebach2024reactive,Lei2025machine} 
    & r\textsuperscript{2}SCAN-D4 
    & 14.31 & 0.07 \\
\bottomrule
\end{tabular}
\vspace{4pt}

\noindent{\footnotesize
$^{a}$~D3 dispersion correction applied via the TorchDFTD3~\cite{takamoto2022towards} calculator
at inference time; the underlying MACE-MP0 model is unchanged.\quad
$^{b}$~All energies referenced to the minimum-energy structure
calculated using DFT and each MLIP; see Figs.~S4--S5 for parity plots.
}
\end{table*}

We apply the proposed reweighting scheme to obtain PMFs for a suite of MLIPs released in 2025. To obtain an estimate of how well the target MLIPs model the studied system, we compared the MLIP-predicted forces and energies with PBE+D3 DFT calculations that are typically used to model nanoporous materials\cite{ferri2026high,sriram2025open}. The reference DFT data, for a subset of 1100 structures sampled along the CV, was obtained from our previous work\cite{majumdar2026free}.
Table~\ref{tab:mlips} shows that all the studied MLIPs forces and energies predictions appear to agree reasonably well with the DFT forces and energies. In particular, we find UMA and ORB to be in excellent agreement with DFT, with force mean absolute errors (MAEs) of forces being 1.81 and 2.72 \si{\kilo\joule\per\mole\per\angstrom} respectively, and MAEs of energies being 0.02 and 0.04 \si{\kilo\joule\per\mole\per\atom} respectively (see Figures S4 and S5 for the parity plots for the entire suite of MLIPs used in this work). 
The MACE-MP0, UMA, and ORB models used in this work have all been trained on PBE-based datasets.
Whereas, MACE-MATPES and MACE-fine-tuned have been trained on different exchange-correlation functionals and consequently exhibit comparatively high force MAEs of 15.47 and 14.31 \si{\kilo\joule\per\mole\per\angstrom}, respectively. 
While MACE-MATPES has been trained on the MATPES dataset with r\textsuperscript{2}SCAN labels\cite{kaplan2025foundational}, MACE-fine-tuned is a fine-tuned version of the MACE-MATPES MLIP developed in this work, tailored for silicate and zeolite systems.
It was trained on a zeolite-based dataset\cite{Erlebach2024reactive,Lei2025machine} with r\textsuperscript{2}SCAN-D4 reference energies and forces (see SI Section~S2 for more details on the training process).

\begin{figure}
    \centering
    \includegraphics[width=\linewidth]{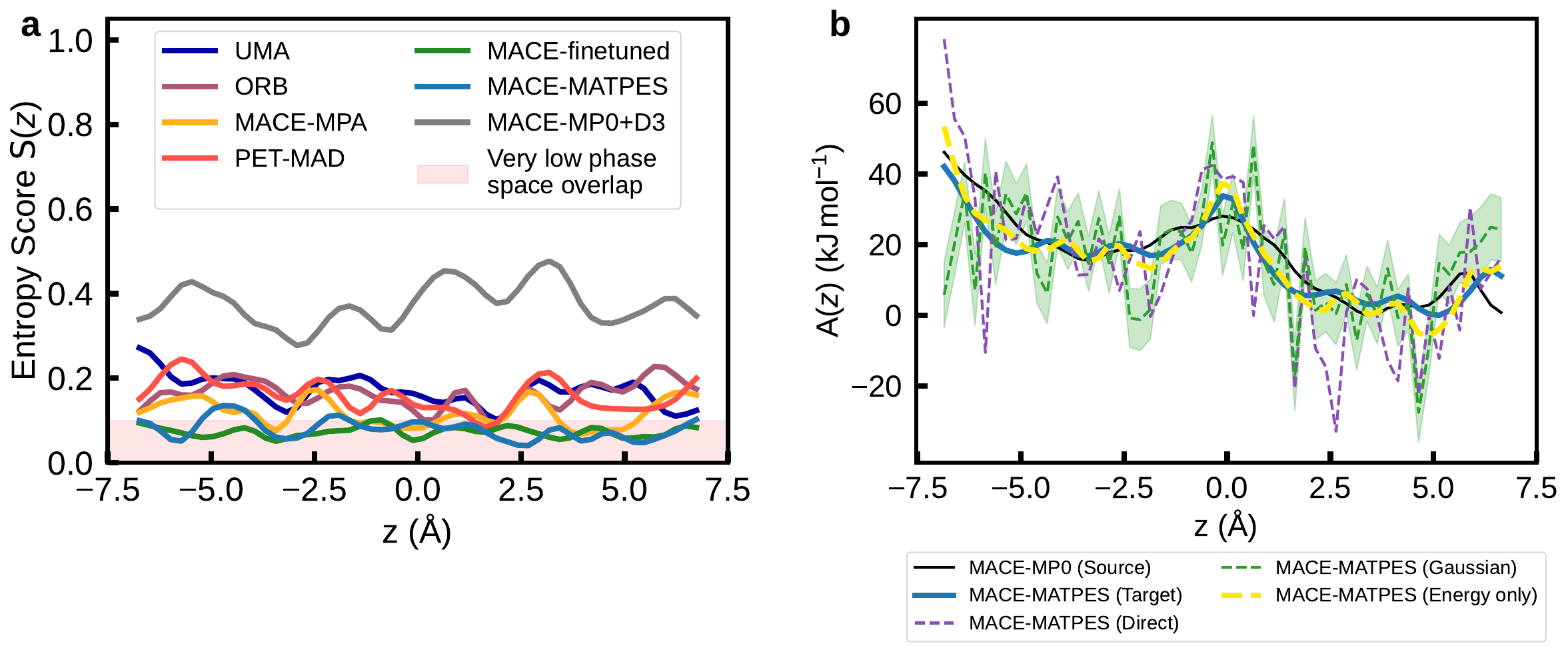}
    \caption{(a) Reweighting entropy score ${S}(z)$ for all target MLIPs, quantifying the effective overlap between the source (MACE-MP0) and target energy distributions at each umbrella window; values near unity indicate very high-phase space overlap or well-distributed (maximum entropy) reweighting weights, while values closer to 0 indicate very low phase space overlap (minimum entropy). The entropy values have been smoothed with a Gaussian filter to eliminate statistical noise and put more emphasis on the trend. (b) PMF obtained from umbrella sampling with MACE-MP0\cite{majumdar2026free} (source) compared with reweighted PMFs based on the MACE-MATPES potential using three methods: direct reweighting, reweighting with Gaussian approximation, and reweighting with energy-only approximation; the shaded areas indicate twice the standard error (see eqn.~(6) in SI). The brute-force umbrella sampling PMF with MACE-MATPES is shown as a reference (target). }
    \label{fig:entropy_matpes}
\end{figure}

Subsequently, we estimate the reweighting entropy score (${S}$) for all the target MLIPs (Figure~\ref{fig:entropy_matpes}a), calculated with eqn.~\eqref{eq:reweighting_entropy}. 
A score of ${S}=1$ corresponds to a uniform weight distribution where every frame in the respective bin along the CV range carries equal reweighting probability (perfect source-target MLIP overlap). 
The entropy approaches 0 when a tiny number of frames dominate the reweighting process. 
MACE-MP0-D3 has the highest entropy score per bin along the CV range (0.4-0.5). 
We find MACE-MATPES and MACE-fine-tuned to have the least overlap (${S}<0.1$) with the source MP0 trajectory. 
The decrease in ${S}$ broadly correlates with the increase in deviation of the MLIP from PBE accuracy (Table~\ref{tab:mlips}).
This motivated us to investigate reweighting to one of the MATPES potentials in greater detail, as they represent particularly challenging reweighting cases, and the solutions developed for them can be broadly applied to the other target MLIPs as well.

Figure~\ref{fig:entropy_matpes}b presents the PMFs reweighted from MACE-MP0 to MACE-MATPES using three distinct estimators compared against the true target PMF (blue curve) obtained from a complete set of umbrella sampling simulations with MACE-MATPES. 
As anticipated, for a system of this size, the exact direct exponential reweighting (purple curve, see section~\ref{sec:reweighting_MBAR} for derivation) exhibits severe statistical noise and fails to converge to the true target PMF. 
This is a consequence of phase-space mismatch between the source and target MLIP; as highlighted by the low reweighting entropy across the entire CV range for MP0$\to$MATPES shown in Figure~\ref{fig:entropy_matpes}a. 
Because the direct free energy reweighting relies on the exponentiated energy gap (Section~\ref{sec:reweighting_MBAR}), and the potential energy difference distribution has such a high variance, it is impossible to faithfully estimate the free energy difference with a reasonable number of samples.
For an in-depth discussion, see SI Section~S6.

To still achieve free energy reweighting, we employ reasonable approximations. 
First, we note that the distribution of potential energy differences between the source and target potentials closely approximates a Gaussian distribution (see Figure~S6-S9). 
This allows us to compute the free energy correction analytically using a closed-form expression as outlined in Section~\ref{sec:gaussian}. 
While this Gaussian approximation (green curve) stabilizes the estimator and partially smooths the recovered PMF, the error in exactly estimating the variance of the energy difference remains sufficiently large to introduce unphysical noise (See Figure~S10 for the Gaussian-approximated PMF of all the target PMFs). 
The target PMF lies within the uncertainty band for almost every bin, confirming that the method is unbiased and might be improved with many more samples, but convergence is expected to be slow.
However, if we take the mathematical limit of ignoring the entropic fluctuations and strictly evaluate the mean energy difference between the source and target MLIPs (Section~\ref{sec:energy_only}), we obtain the energy-only reweighted PMF (Figure~\ref{fig:entropy_matpes} b, yellow curve). 
This approximation successfully recovers a PMF that is very similar to the explicitly sampled, brute-force MACE-MATPES PMF.
The energy-only MP0$\to$MATPES reweighted PMF shows the three local minima per cavity as the target PMF, which are not absent in the MP0 source PMF. 
Furthermore, the barrier separating the two cavities has become much narrower, as is the case for the MATPES PMF.
Hence, even though the severe energy-only approximation cannot yield perfect agreement, it does reproduces all the structural details of the target PMF that are absent in the source one.

\begin{figure}[htbp]
    \centering
    \includegraphics[width=0.9\linewidth]{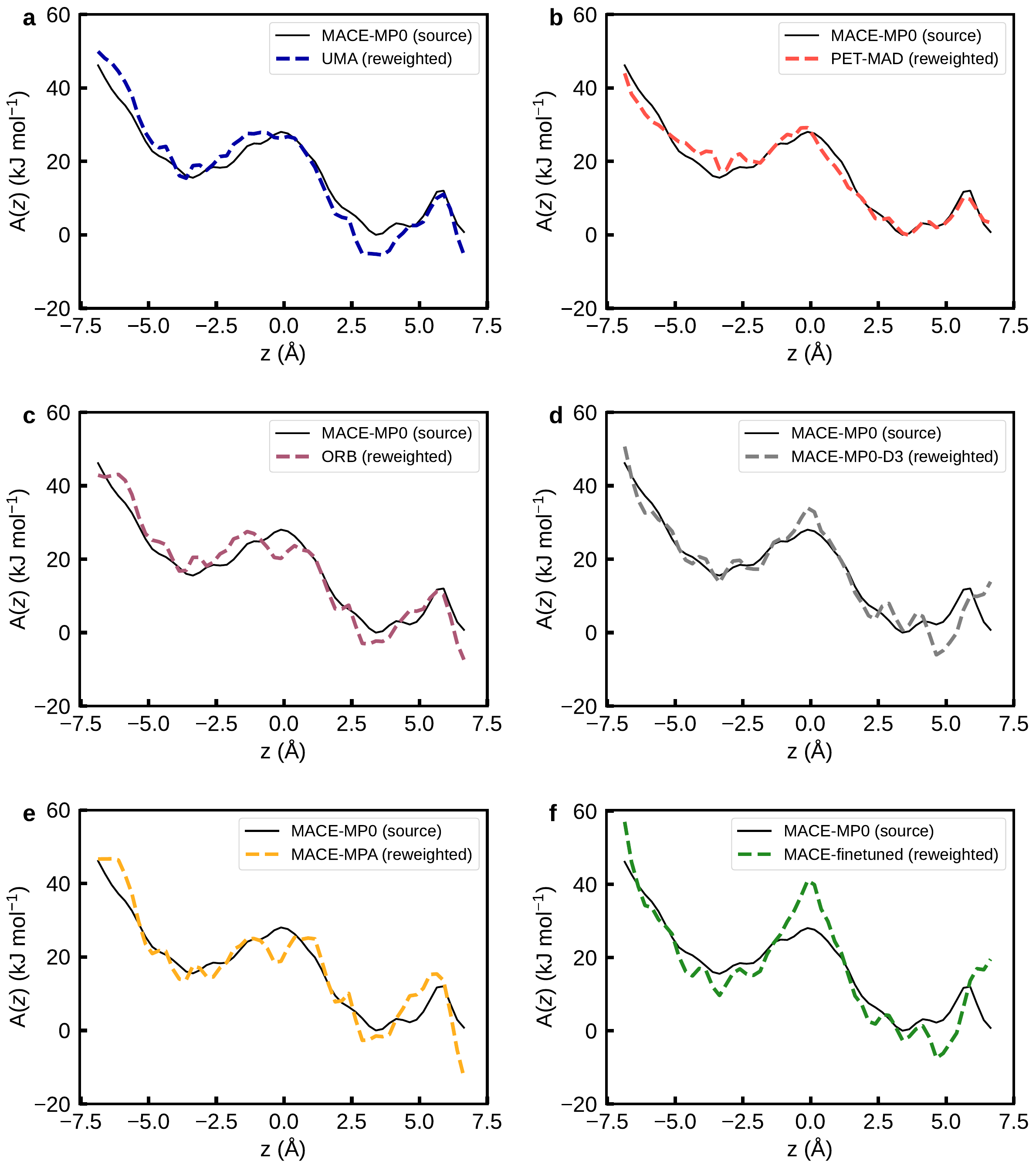}
    \caption{(a–f) Reweighted PMFs obtained using the energy-only approximation (Eq.~(\ref{eq:energy_only})) for six MLIPs, (a) UMA, (b) PET-MAD, (c) ORB, (d) MACE-MP0-D3, (e) MACE-MPA, and (f) MACE-fine-tuned, starting from the MACE-MP0 umbrella sampling trajectory (black). Dashed lines show the reweighted PMF for each target MLIP.}
    \label{fig:all_mlips}
\end{figure}

Having established that the energy-only estimator produces a reasonably accurate reweighted PMF for MACE-MATPES, which is one of the MLIPs with the lowest reweighting entropy score for the source MACE-MP0 potential and therefore the stress test for our approach, we apply this methodology to the remaining target MLIPs and obtain the respective reweighted PMFs, as shown in Figure~\ref{fig:all_mlips}. 
From a visual or qualitative inspection, we find UMA, PET-MAD, ORB, MACE-MP0-D3, and MACE-MPA reweighted PMFs to be similar to each other and the source MACE-MP0 PMF. The MP0$\to$MACE-fine-tuned  PMF, on the other hand, appears to be less similar to the source and the other reweighted PMFs. Trained with higher fidelity DFT data (r\textsuperscript{2}SCAN-D4), MACE-fine-tuned, also like MACE-MATPES, has a comparatively low phase space overlap with MACE-MP0. Surprisingly, ORB and MACE-MPA-reweighted PMFs exhibit a local minimum at the transition state ($z\approx 0~\text{\AA}$).

\subsection{Quantifying similarity between reweighted PMFs}\label{sec:quantifying}

\begin{figure}[H]
    \centering
    \includegraphics[width=\linewidth]{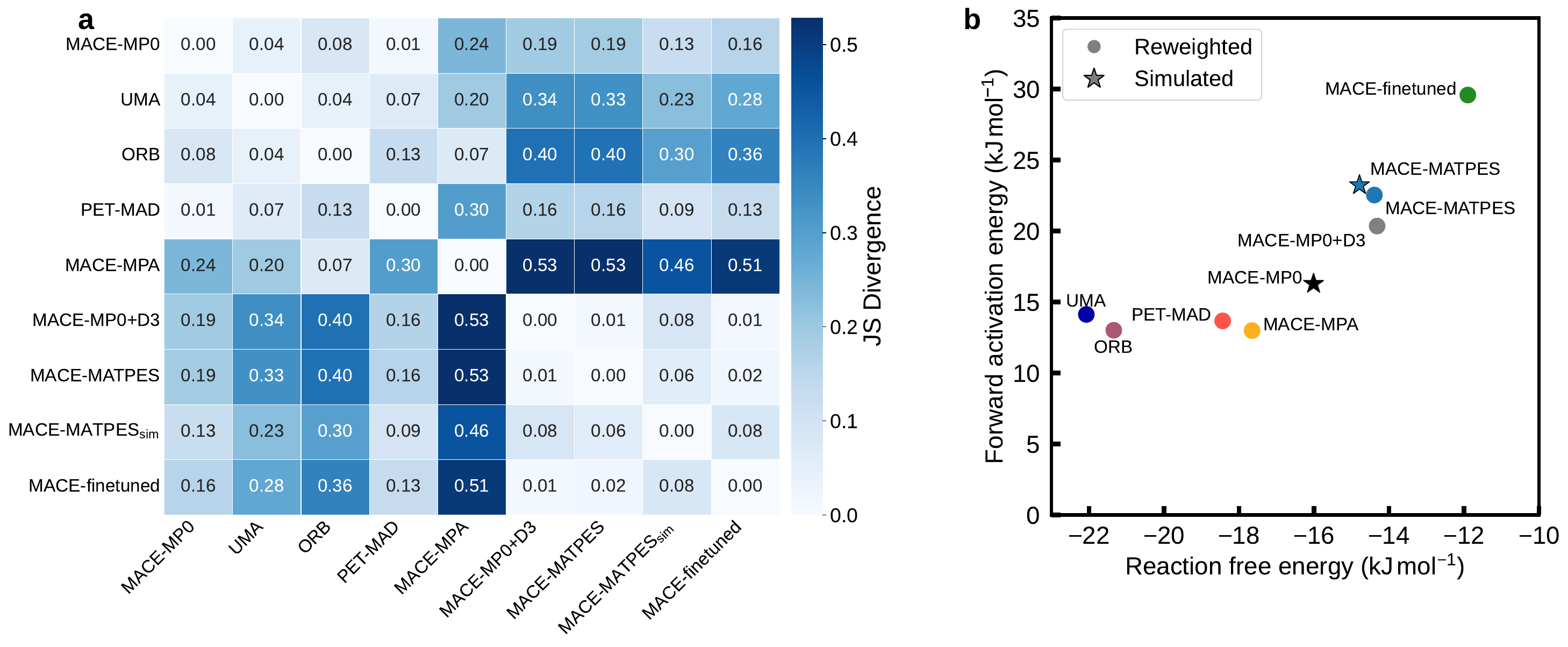}
    \caption{(a) Pairwise Jensen--Shannon (JS) divergence between the PMFs of different MLIPs obtained with energy-only reweighting; the MACE-MATPES$_\text{sim}$ entry corresponds to the fully simulated PMF. (b) Reaction free energy (x-axis) and forward activation free energy barrier (y-axis) computed from the PMFs of the different MLIPs.}
    \label{fig:metrics}
\end{figure}

To quantify the similarity between the reweighted PMFs, we computed pairwise Jensen–Shannon divergence (JSD) scores, shown as a heatmap in Figure~\ref{fig:metrics}a. Lighter colors indicate greater similarity between a given pair of PMFs. Several notable patterns emerge. The heatmap reveals two distinct intra-cluster blocks with low JSD values: an upper-left block comprising MACE-MP0, UMA, ORB, and PET-MAD (`cluster 1'), and a lower-right block comprising MACE-MP0-D3, MACE-MATPES, and MACE-fine-tuned (`cluster 2'). When both the source and target MLIPs belong to the same cluster, reweighting between them can be carried out with very high confidence. This intra-cluster similarity largely reflects the shared characteristics of the datasets and exchange-correlation functionals used to train the respective  MLIPs (see Table~\ref{tab:mlips}). In contrast, the inter-cluster blocks, corresponding to source-target pairs drawn from different clusters (\textit{e.g.}, MACE-MP0 as source and MACE-MATPES as target), exhibit higher JSD values, indicating very low phase-space overlap owing to the difference in the respective underlying training data. Interestingly, MACE-MPA occupies an intermediate position: its JSD values with `cluster 1' MLIPs, particularly ORB, are lower than those with `cluster 2' MLIPs, suggesting it is more closely related to `cluster 1' in terms of the underlying phase-space distribution.

For practical applications, chemists and materials scientists are often interested in thermodynamic properties such as the activation energy and the reaction free energy of the process under study. Figure~\ref{fig:metrics} b presents these two quantities for all considered MLIPs, computed following the process outlined in Refs.~\cite{dietschreit2022free,dietschreit2022obtain}. To avoid the accumulation of errors in reweighting across all bins along the CV, we split the CV axis into three regions (reactant, transition state, and product) and apply the energy-only reweighting technique to each region as a whole. This `3-split' approach avoids bin-level noise
while retaining the distinction between the three physically meaningful states.  We find that the simulated and reweighted MACE-MATPES properties are in excellent agreement with each other (within ${\approx}4$ \si{kJ/mol}, i.e., chemical accuracy), thereby validating the success of the demonstrated reweighting strategy in this work.  Furthermore, an interesting pattern emerges that echoes the JSD clustering in panel (a): `cluster 1' MLIPs exhibit similar forward activation energy in the range ${\approx}14$ to $16$ \si{kJ/mol}, while cluster 2 MLIPs yield similar reaction free energies in the range ${\approx}-12$ to $-15$ \si{kJ/mol}. MACE-MPA aligns more closely with `cluster 1' MLIPs, similar to the findings from JSD values. The source MLIP, MACE-MP0, lies at the intersection of these two groups: its activation free energy aligns with `cluster 1', while its reaction free energy falls within the range of `cluster 2', thereby occupying a thermodynamic middle ground between the two classes of MLIPs.

This behavioral split underscores the fact that the reference exchange-correlation functional (\textit{e.g.}, PBE, PBE+D3, PBE-sol, r\textsuperscript{2}SCAN, r\textsuperscript{2}SCAN-D4), and the underlying training datasets (\textit{e.g.}, MP-Trj, Alexandria, MATPES, OMAT, MAD) (see Table~\ref{tab:mlips}) together dictate the fundamental topology of the free energy landscape. The choice of the neural network architecture, such as using an equivariant model (MACE-MP0, UMA) versus a non-equivariant one (PET-MAD, ORB), does not have a significant impact on the reweighted PMFs and derived thermodynamic quantities.
Ultimately, the analysis presented here provides a robust framework for determining in which cases reweighting can be carried out without the necessity of brute force umbrella sampling simulations (such as in the case of intra-cluster reweighting). Additionally, reweighting from a lower (MACE-MP0, UMA, ORB) to a higher (MACE-MATPES, MACE-fine-tuned) fidelity MLIP, a fundamentally challenging task, can also be carried out with high success, following the methodology outlined with the energy-only estimator and demonstrated for the MACE-MATPES MLIP. However, we note that the results with this approximation should be carefully interpreted as a lower bound, and can benefit from verification with additional umbrella sampling simulations as shown in Figure~\ref{fig:entropy_matpes} b.

\section{Conclusions and outlook}
In this work, we presented a systematic and highly scalable framework for reweighting potentials of mean force (PMFs) obtained from a single `source' MLIP to a representative suite of `target' MLIPs (see Table.~\ref{tab:mlips}).
Outlining the fundamental limitations of direct reweighting, which fails for system sizes of practical interest due to the low phase-space overlap between MLIPs trained on different datasets and exchange-correlation functionals, we demonstrated that analytical corrections, such as Gaussian-based and mean-energy-difference-based approximations, can successfully bypass these limitations. Our results show that the latter can provide a reasonably accurate estimate of the target PMF. This approach enabled us to evaluate the thermodynamics of \ce{Li+} transport in nanoconfinement (a complex 601-atom water-solvated zeolite system) at different levels of \textit{ab-initio} accuracy (PBE-D3, PBE-sol, r\textsuperscript{2}SCAN, r\textsuperscript{2}SCAN-D4) and at a mere fraction of the computational cost required for entire umbrella sampling simulations. 

To quantify the similarity between MLIP-reweighted PMFs, we further calculated Jensen-Shannon Divergence (JSD) scores and macroscopic thermodynamic properties of interest, namely reaction and activation free energies. This analysis revealed that the studied MLIPs partition themselves into two physically distinct clusters: one trained on PBE and related intra-GGA-based datasets and the other trained on r\textsuperscript{2}SCAN and/or dispersion (D3/D4)-corrected data. To obtain PMFs and derived thermodynamic properties with a target MLIP that falls in the same cluster as the source MLIP, reweighting using the mean energy difference estimator is likely to be highly successful, without the necessity of performing expensive simulations. 

More importantly, the practical utility of the proposed reweighting strategy extends even to highly divergent models. The thermodynamic properties derived from the MACE-MATPES reweighted PMFs were found to be in excellent agreement with the corresponding values obtained from explicit umbrella sampling. The fact that this strategy yielded such accurate consensus for a target MLIP exhibiting very low phase-space overlap ($S<0.1$) with the source MLIP underscores its stability and scalability to system sizes of several hundreds of atoms. Thus, when translating across distinct clusters, reweighting with mean-energy approximation can serve as a highly efficient lower-bound estimate that can be verified with enhanced sampling if absolute chemical precision is required.

Overall, this work establishes robust guidelines for determining the utility and also the limits of MLIP-based thermodynamic reweighting. As the ecosystem of MLIPs continues to rapidly expand, this work provides an important diagnostic protocol. Rather than performing high-resource sampling for every emerging model, researchers can utilize the reweighting strategy discussed in this work to obtain rapid cross-model consensus on materials chemistry properties for system sizes of practical interest, effectively addressing a major bottleneck in computational materials design.

\section{Methods}\label{sec:methods}

\subsection{Umbrella Sampling}

The umbrella sampling simulations using the MACE-MATPES potential were performed at 450~K with the NFF package\cite{nff_github}, following the same procedure as outlined in our previous work\cite{majumdar2026free}.

\subsection{Reweighting methodology}
\subsubsection{Direct reweighting with MBAR}\label{sec:reweighting_MBAR}

For the umbrella simulations, we chose the bias energy $U^{\text{bias}}_i(\mathbf{x})$ of a configuration $\mathbf{x} = (x_1, x_2, \dots, x_{3N_a})^\top$, with $N_a$ being the number of atoms to take the form:
\begin{equation}\label{eq:direct_reweighting}
    U^\mathrm{bias}_i(\mathbf{x})=\frac{k_i}{2}(\xi(\mathbf{x})-z_i)^2 \ ,
\end{equation}
where $k_i$ is the spring constant and $z_i$ is the equilibrium value for 
umbrella window $i$.
$\xi(\mathbf{x})$ denotes the collective variable, which maps the high-dimensional configuration onto a single scalar.

To recover the unbiased statistics, we used MBAR\cite{shirts2008statistically} to estimate the  relative, dimensionless free energies $f_i$ introduced by each biasing  potential $U^\mathrm{bias}_i$, which are defined implicitly through
\begin{equation}
	f_i = - \ln 
	\sum_k^K \sum^{N_k}_j 
	\frac{e^{-\beta U^\mathrm{bias}_i(k,j)}}
	{\sum_l^K N_l\, e^{f_l - \beta U^\mathrm{bias}_l(k,j)}} \ ,
	\label{eq:MBAR} 
\end{equation}
where $K$ is the number of windows, $N_i$ is the number of frames in  window $i$, and $U^\mathrm{bias}_i(k,j)$ is the value of the biasing potential of window $i$ evaluated at frame $j$ drawn from simulation window $k$. 
As usual, $\beta = 1 / k_\mathrm{B}T$, with $k_\mathrm{B}$ being the Boltzmann constant and $T$ the absolute temperature.
In this work, $K=54$ and for every window $k$, $N_k=10,000$.

The unbiased Boltzmann weight of a configuration $\mathbf{x}$ is  then
\begin{equation}
	W(\mathbf{x}) = \frac{1}{\sum_k^K N_k\, e^{f_k - \beta U^\mathrm{bias}_k(\mathbf{x})}} \ .
\end{equation}
These weights can be used to build the probability density $\rho(z)$ as a histogram along the collective variable,
\begin{equation}
	\rho(z) = \bigl\langle \delta(\xi(\mathbf{x}) - z)\bigr\rangle = \sum_k^K \sum_j^{N_k} W(k,j)\, \delta\bigl(\xi(k,j) - z\bigr) 
	 \ ,
\end{equation}
where $\delta(\dots)$ is the Dirac delta function that assigns each frame to a bin along the CV axis.
From here, the potential of mean force (PMF) follows as
\begin{equation}
	A(z) = - k_\mathrm{B}T \ln \rho(z) \ .
\end{equation}

If we now seek the PMF for a different PES $U_\mathrm{B}$ without running new, expensive simulations, one can reweight the frames already sampled under $U_\mathrm{A}$:
\begin{equation}
	W_\mathrm{rw}(\mathbf{x}) 
	=  \frac{e^{-\beta \left(U_\mathrm{B}(\mathbf{x}) - U_\mathrm{A}(\mathbf{x})\right)}}
	{\sum_k^K N_k\, e^{f_k - \beta U^\mathrm{bias}_k(\mathbf{x})}} 
	= e^{-\beta \Delta U(\mathbf{x})}\, W(\mathbf{x}) \ ,
    \label{eq:correcting_weights}
\end{equation}
where $\Delta U = U_\mathrm{B} - U_\mathrm{A}$. 
The PMF for $U_\mathrm{B}$ is then
\begin{equation}\label{eqn:PMF}
	A_\mathrm{B}(z) = - k_\mathrm{B}T\, \ln \langle \delta(\xi(\mathbf{x}) - z) \rangle_\mathrm{B} = - k_\mathrm{B}T\, \ln 
	\sum_k^K \sum_j^{N_k} W_\mathrm{rw}(k,j)\, \delta\bigl(\xi(k,j) - z\bigr) \ .
\end{equation}

\subsubsection{Reweighting entropy}\label{sec:entropy}

For a specific subset of $N$ frames assigned to a specific bin along the reaction coordinate $z$, we define the normalized statistical probability $p_z(i)$ of each frame $i$ as:
\begin{equation}
    p_z(i) = \frac{W_\mathrm{rw}(i)}{\displaystyle\sum_{j=1}^{N} W_\mathrm{rw}(j)} \ .
\end{equation}
using the reweighted Boltzmann weights $W_\mathrm{rw}$, as calculated with eqn.~\eqref{eq:correcting_weights}.
The reweighting entropy score is then defined as the Shannon entropy of this discrete probability distribution $\{p(i)\}$, strictly normalized by its theoretical maximum ($\ln N$):
\begin{equation}\label{eq:reweighting_entropy}
    {S}(z) = \frac{-\displaystyle\sum_{i=1}^{N} p_z(i)\ln p_z(i)}{\ln N} \;\in\; [0,\,1] \ .
\end{equation}

\subsubsection{Gaussian approximation for PMF reweighting}\label{sec:gaussian}

Rather than evaluating the target PMF directly, it is more instructive to express $A_\mathrm{B}(z)$ as a correction to the original reference PMF, $A_\mathrm{A}(z)$. 
Using the identity $\langle O \rangle_\mathrm{B} \propto \langle O e^{-\beta\Delta U} \rangle_\mathrm{A}$ and factoring out the conditional average at fixed $\xi=z$, the target PMF from eqn.~\eqref{eqn:PMF} can be expanded as:
\begin{equation}\label{eq:pmf_correction}
\begin{split}
    A_\mathrm{B}(z) &= -k_B T \ln \left( \langle \delta(\xi(\mathbf{x}) - z) \rangle_\mathrm{A} \langle e^{-\beta\Delta U} \rangle_{\mathrm{A},\xi=z} \right) + C \\
    &= A_\mathrm{A}(z) - k_B T \ln \langle e^{-\beta\Delta U} \rangle_{\mathrm{A},\xi=z} + C \ ,
\end{split}
\end{equation}
where $\Delta U = U_\mathrm{B} - U_\mathrm{A}$ is the potential energy difference, $\langle \dots \rangle_{\mathrm{A},\xi=z}$ denotes the ensemble average over the reference potential $U_\mathrm{A}$ strictly restricted to frames falling within the bin centered on $z$, and $C$ is an additive constant independent of $z$ that can be safely absorbed into the overall normalization.

Because the direct evaluation of the conditional average $\langle e^{-\beta\Delta U} \rangle_{\mathrm{A},\xi=z}$ is notoriously ill-conditioned, and the distribution of energy differences sampled under $U_\mathrm{A}$ is reasonably close to a Gaussian (Figure~S5), a more stable route is to compute the correction to the PMF analytically. Substituting the Gaussian approximation into the conditional average gives a closed-form result:
\begin{equation}\label{eq:gaussian_integral}
\begin{split}
    \langle e^{-\beta\Delta U}\rangle_{\mathrm{A},\xi=z} &\approx \int_{-\infty}^{\infty} d(\beta\Delta U) \mathcal{N}(\beta\Delta U;\mu_z,\sigma_z) e^{-\beta\Delta U} \\
    &= e^{-(2\mu_z-\sigma_z^{2})/2} \int_{-\infty}^{\infty} d(\beta\Delta U) \mathcal{N}(\beta\Delta U;\mu_z-\sigma_z^{2},\sigma_z) \\
    &= e^{-(2\mu_z-\sigma_z^{2})/2} \ ,
\end{split}
\end{equation}
where the last step uses the fact that a Gaussian integrates to unity. 
Here, $\mu_{z}$ and $\sigma_{z}$ are the mean and standard deviation of the dimensionless energy difference $\beta\Delta U$ fitted within the bin $\xi\approx z$.

The PMF correction then takes the simple form:
\begin{equation}\label{eq:gaussian_pmf}
\begin{split}
    A_\mathrm{B}(z) - A_\mathrm{A}(z) &= -k_B T \ln \langle e^{-\beta\Delta U}\rangle_{\mathrm{A},\xi=z} \\
    &= k_B T\, \frac{2\mu_{z}-\sigma_{z}^{2}}{2}.
\end{split}
\end{equation}
This approach is numerically better conditioned than the direct frame-by-frame procedure because it sidesteps the explicit evaluation of the exponential average.

\subsubsection{Energy-only approximation for PMF reweighting}\label{sec:energy_only}

If the energy gap distributions deviate too far from a Gaussian for the standard deviation alone to characterize them faithfully or if the estimated variance $\sigma_{z}^{2}$ cannot be estimated with sufficient accuracy, one can make a cruder but highly stable approximation by discarding the shape of the distribution entirely. 
Formally, this amounts to neglecting the entropic contribution to the free energy difference. 

By retaining only the mean energy gap and ignoring the width of the fluctuations, the PMF correction simplifies to the mean-energy, or energy-only, approximation:
\begin{equation}\label{eq:energy_only}
    A_\mathrm{B}(z) \approx A_\mathrm{A}(z) + \langle \Delta U \rangle_{\mathrm{A},\xi=z}.
\end{equation}

This is a much simpler correction and should be treated as a lower bound, since it will systematically underestimate the correction whenever the $\Delta U$ distribution is broad. However, by completely bypassing the statistical noise inherent to the exponential average and the variance terms, this approximation results in a substantially smoother and more converged estimate of the target PMF.

\subsection{Jensen--Shannon divergence}
The JS divergence $D_{\text{JS}}$ between two probability distributions $A$ and $B$ is given by 
\begin{equation}
    D_\text{JS}(A \parallel B)=\frac{1}{2}(D_{\text{KL}}(A \parallel M)+ D_{\text{KL}}(B \parallel M)),
\end{equation}
where $M=\frac{1}{2}(A+B)$ and $D_{\text{KL}}$ is the Kullback--Leibler divergence that is defined as

\begin{equation}
    D_{\text{KL}}(A\parallel B)=\sum _{x\in {\mathcal {X}}}A(x)\,\ln {\frac {A(x)}{B(x)}}
\end{equation}
over the sample space $\mathcal{X}$.
In this work the JS divergence has been computed with the scipy.spatial.distance module\cite{2020SciPy-NMeth}.

\section{Code and data availability}
The code associated with this work are provided in the GitHub repository\cite {reweighting_github}. The data files associated with this work are provided in the Zenodo repository dedicated to this work\cite{majumdar_2026_20142456}.

\section{Acknowledgement}

S.M.~is supported by the Postdoctoral Mobility Fellowship from the Swiss National Science Foundation (SNSF), grant number P500PN\textunderscore222184. M.S.~gratefully acknowledges the Postdoctoral Mobility fellowship P500PN\textunderscore225736 from the Swiss National Science Foundation. L.G.~and D.W.~acknowledge the support of the Czech Science Foundation(26-23277S), Ministry of Education, Youth and Sports of the Czech Republic through the e-INFRA CZ (ID:90254), and MIT-Czech Republic IOCB Tech Foundation Seed Fund.
The authors acknowledge the MIT Engaging cluster and MIT Lincoln Laboratory Supercloud cluster at the Massachusetts Green High-Performance Computing Center (MGHPCC) for providing high-performance computing resources\cite{reuther2018interactive}.

\section{Supporting Information available}
The Supporting Information includes: visualization of the studied system; a short note on the developed MACE-fine-tuned; efficacy of downsampled frames; parity plots; additional free energy reweighting results; standard error estimates; further background on reweighting and its limits; the following additional references were cited in the SI\cite{FPS,SOAP}.

\section{Author contributions}
S.M. and R.G.-B. conceived the project. S.M. developed the reweighting computational workflow, performed umbrella sampling simulations, computed MLIP forces and energy, thermodynamic properties, and led the manuscript drafting. M.S. performed the framewise energy difference analysis and integrated the PET-MAD MLIP calculations into the reweighting workflow. J.C.B.D. analyzed the limits of reweighting and provided the mathematical framework. J.C.B.D., M.S., and S.R. contributed to the development of the reweighting workflow, testing, and provided feedback on improvement. D.W., L.G., and S.R. developed the MACE-fine-tuned model. J.C.B.D, and R.G.-B. provided supervision and strategic guidance.  All authors discussed the results and contributed to the scientific narrative of the manuscript.

\bibliography{sn-bibliography}

\end{document}



\title[Article Title]{Supporting Information: Reweighting free energy profiles between universal machine learning interatomic potentials for fast consensus building}

\author[1]{\fnm{Sauradeep} \sur{Majumdar}} 
\author[1]{\fnm{Miguel} \sur{Steiner}} 
\author*[2]{\fnm{Johannes C. B.} \sur{Dietschreit}\email{johannes.dietschreit@univie.ac.at}}
\author[1]{\fnm{Swagata} \sur{Roy}}
\author[3]{\fnm{Daniel} \sur{Willimetz}}
\author[3]{\fnm{Lukaš} \sur{Grajciar}}
\author*[1]{\fnm{Rafael} \sur{G\'omez-Bombarelli}\email{rafagb@mit.edu}}

\affil*[1]{\orgdiv{Department of Material Science and Engineering}, \orgname{Massachusetts Institute of Technology}, \orgaddress{\city{Cambridge}, \state{Massachusetts}, \postcode{02139}, \country{United States}}}
\affil[2]{\orgdiv{Institute of Theoretical Chemistry}, \orgname{University of Vienna}, \orgaddress{\street{Währinger Str. 17}, \postcode{1090}, \city{Vienna}, \country{Austria}}}
\affil[3]{\orgdiv{Department of Physical and Macromolecular Chemistry}, \orgname{Charles University}, \orgaddress{\street{Hlavova 8}, \city{Praha 2}, \postcode{12800}, \country{Czech Republic}}}

\maketitle

\tableofcontents

\section{Visualization of the system studied in this work}
\begin{figure}[ht]
    \centering
    \includegraphics[width=0.7\linewidth]{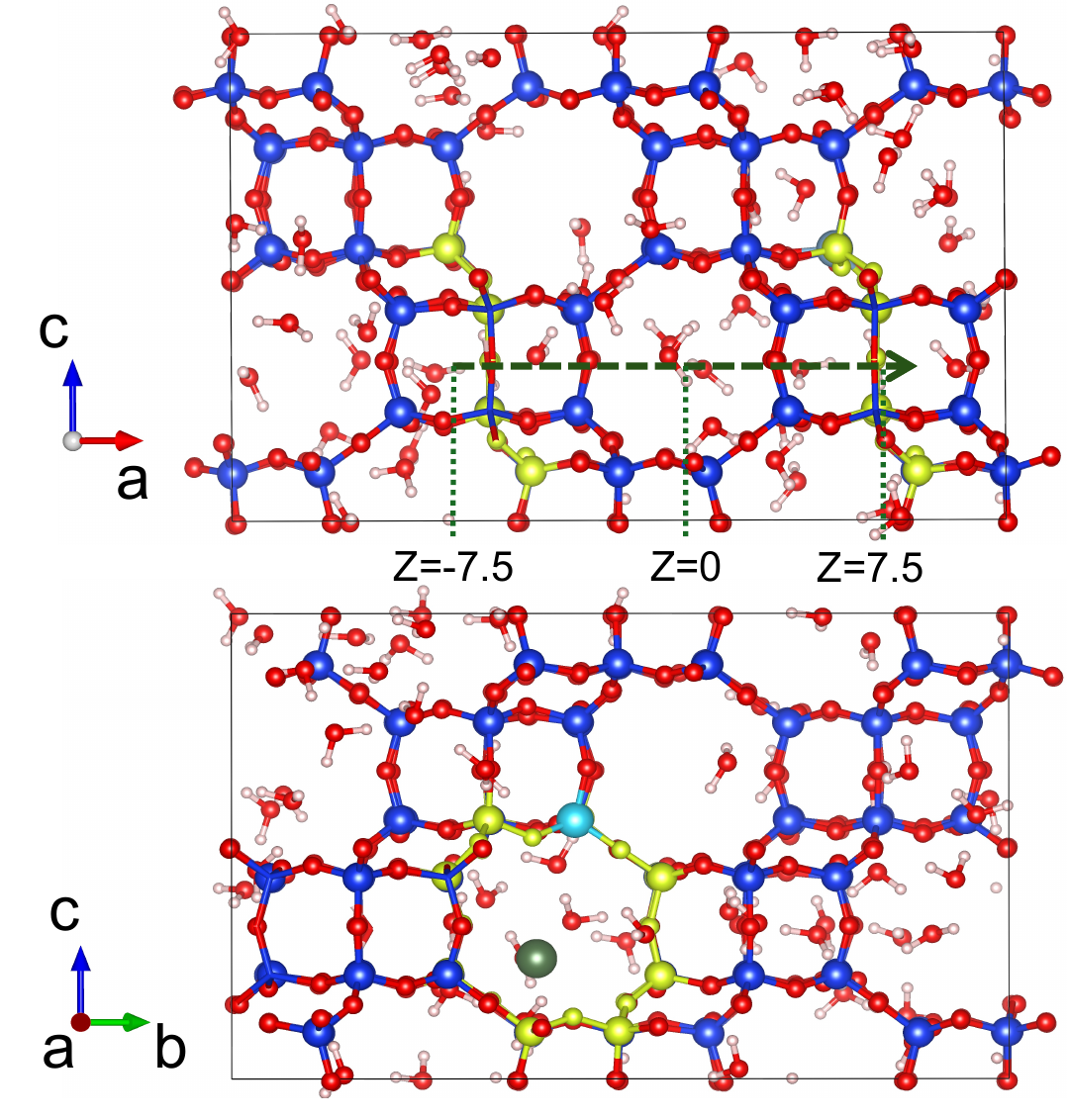}
    \caption{Snapshot of the system, \ce{Li+} diffusion along the pore channel of a CHA zeolite, in two different views. Top: The green dotted arrow highlights the CV(z) used to bias the \ce{Li+} along the pore channel formed by the 8-membered rings (highlighted by yellow atoms) of the zeolite. Bottom: Same system rotated clockwise by 60 degrees, with the pore channel formed by the 8-membered rings pointing into/out of the plane of the paper/screen (denoted by the `a' axis). The \ce{Li+} has been inserted in the center of one of the 8-membered rings, and one of the Si atoms in the ring has been replaced with an Al atom. Color coding of atoms: Red=O, Blue=Si, White=H, Light Blue=Al, Dark Green=Li. Image adapted with permission from Majumdar et al.\cite{majumdar2026free}}
    \label{fig:starting_system}
\end{figure}
\newpage

\section{Short note on the fine-tuned MLIP developed in this work}

In addition to using pre-trained MLIPs, we developed a fine-tuned version of the MACE-MATPES-r2SCAN foundational model\cite{mace_github}, specifically optimized for silicate and zeolite systems. The training set is based on an extensive SCAN-D3 database from Erlebach et al.\cite{Erlebach2024reactive, Lei2025machine}, comprising over 330,000 configurations of siliceous and aluminosilicate zeolites, which has been recalculated with a yet higher level r2SCAN-D4 potential. While the initial training databases provided a robust description of Al, H, Na, O, and Si, using this model for the case study in this work requires an accurate representation of Li environments. To circumvent the high computational cost of generating a large-scale $\omega$B97X-D3 dataset for a new element, we employed a generalizable active-learning-inspired approach leveraging the chemical similarity between Na and Li. We first performed Furthest Point Sampling (FPS) \cite{FPS} on Na-centered SOAP descriptors \cite{SOAP} to isolate 5,000 diverse Na environments from the original database. In this subset, 75\% of Na atoms were replaced with Li, and the systems were equilibrated via 20 ps of molecular dynamics using the MACE-MP0 foundational model. Following convergence verification via the autocorrelation function, the final 5 ps of the trajectories yielded a pool of 500,000 structures, which were further refined via Li-centered SOAP-FPS. These selectively sampled structures were recomputed at the r2SCAN-D4 level to maintain consistency with the reference zeolite database. By retaining the initial 5,000 Na-based structures, we produced a final training set of 10,000 high-diversity configurations, enabling the MACE-fine-tuned model to accurately describe Li-ion diffusion and Na/Li interactions within the zeolite framework. 

The fine-tuned model was trained for 50 epochs using a distance cutoff of 5~\AA, 
128 channels, and a maximum angular momentum ($L_{\mathrm{max}}$) of 1. The resulting model achieved test errors of 
\SI{0.23}{\kilo\joule\per\mole\per\atom} for energies and \SI{3.65}{\kilo\joule\per\mole\per\angstrom} for forces. 
The MACE-MATPES model with a post hoc D4 dispersion correction evaluated on the same test set yielded errors of \SI{1.48}{\kilo\joule\per\mole\per\atom} for energies and \SI{6.22}{\kilo\joule\per\mole\per\angstrom} for forces. These findings demonstrate that fine-tuning the MACE-MATPES foundational model improves predictive accuracy for the targeted application domain.
\newpage

\section{Efficacy of downsampled frames}
\begin{figure}[ht]
    \centering
    \includegraphics[width=0.8\linewidth]{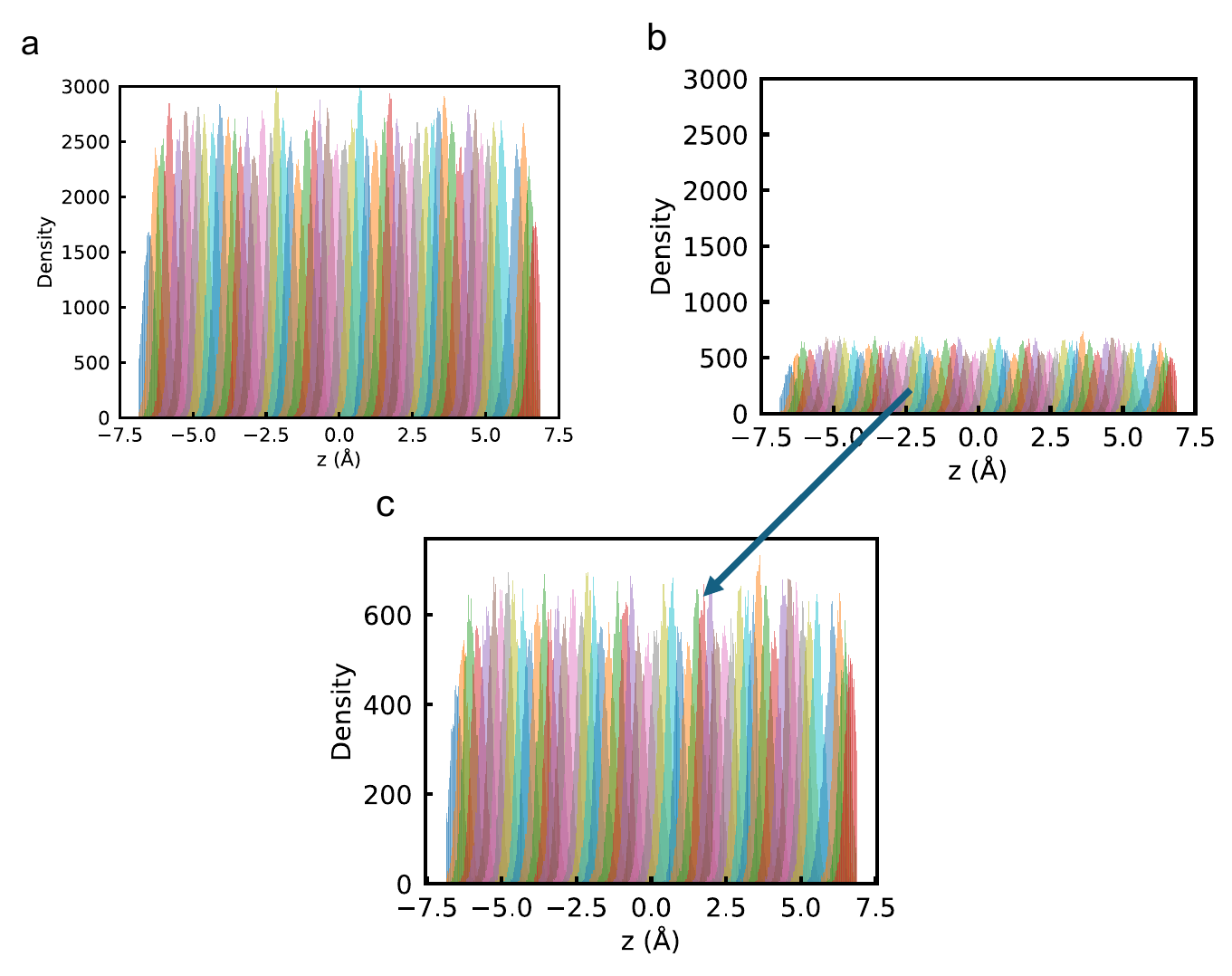}
    \caption{(a) Histogram of umbrella windows along the CV(z), as sampled by the base simulation using MACE-MP0. (b) Histogram of umbrella windows, downsampled along the CV (z), with 10000 structure frames per window. (c) Zoomed-in view of (b). }
    \label{fig:density}
\end{figure}

\begin{figure}[H]
    \centering
    \includegraphics[width=0.6\linewidth]{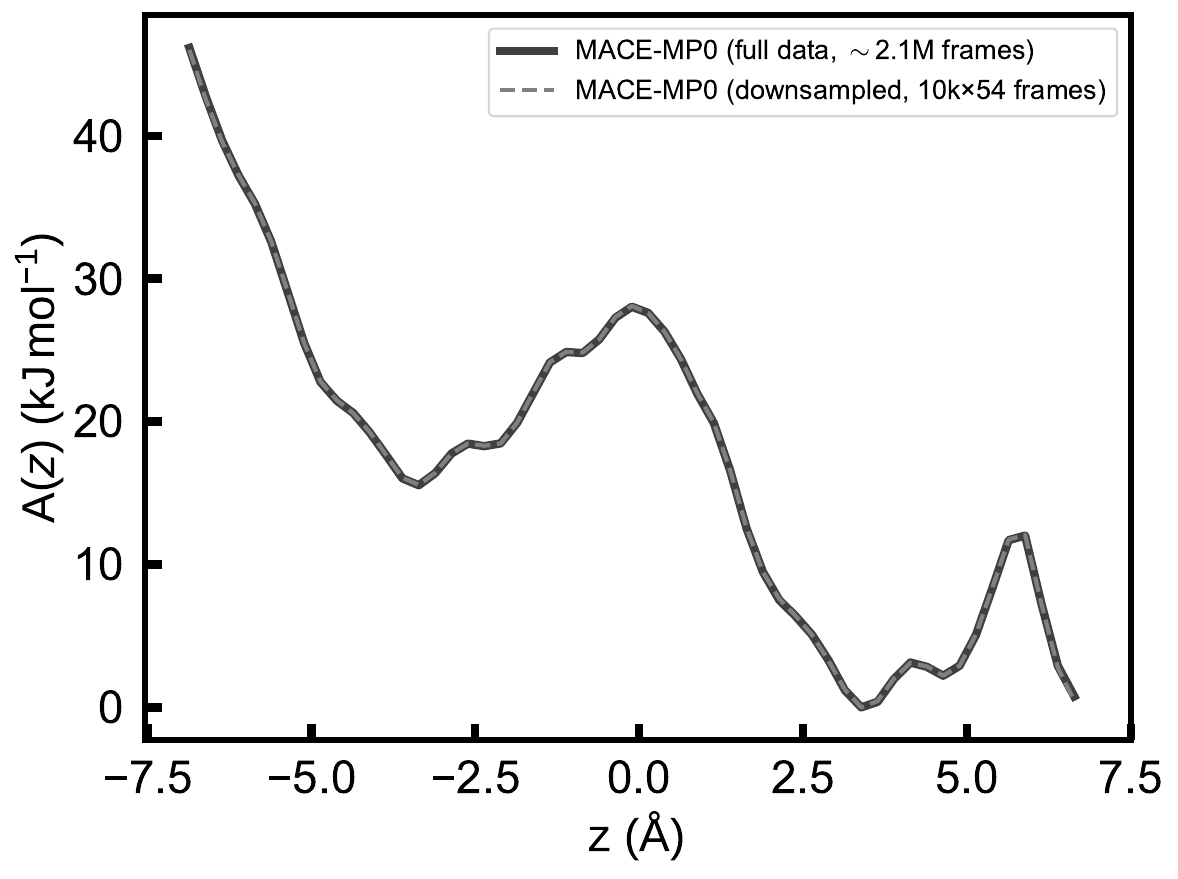}
    \caption{10000 samples diverse downsampling}
    \label{fig:placeholder}
\end{figure}

\newpage
\section{Parity plots}
\begin{figure}[H]
    \centering
    \includegraphics[width=\linewidth]{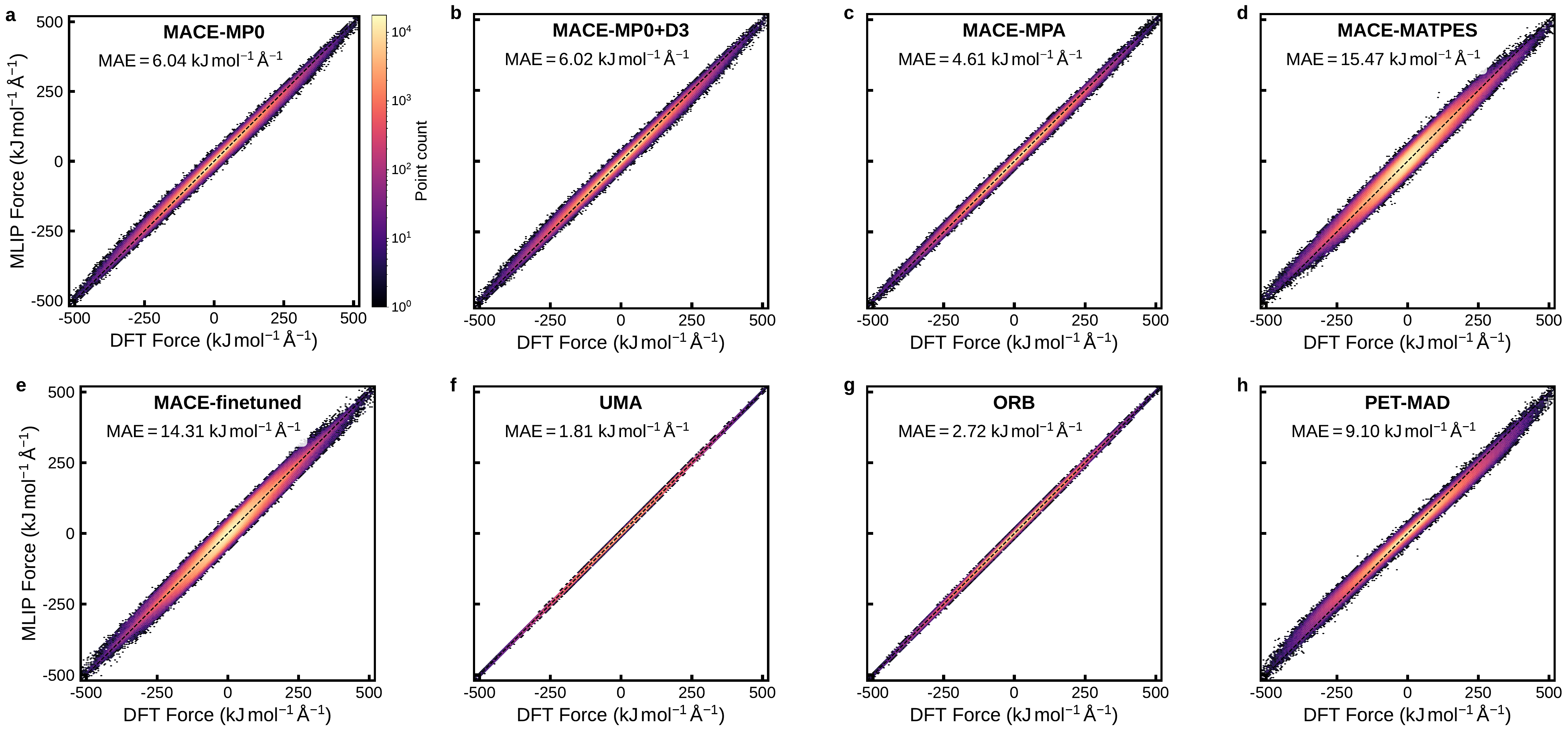}
    \caption{Parity plots for forces of the Li-zeolite-\ce{H2O} system, obtained with DFT (PBE+D3) and the suite of MLIPs used in this work. A total of 1100 structures were used for the parity plots, each with 601 atoms.}
    \label{fig:parity_force}
\end{figure}

\begin{figure}[H]
    \centering
    \includegraphics[width=\linewidth]{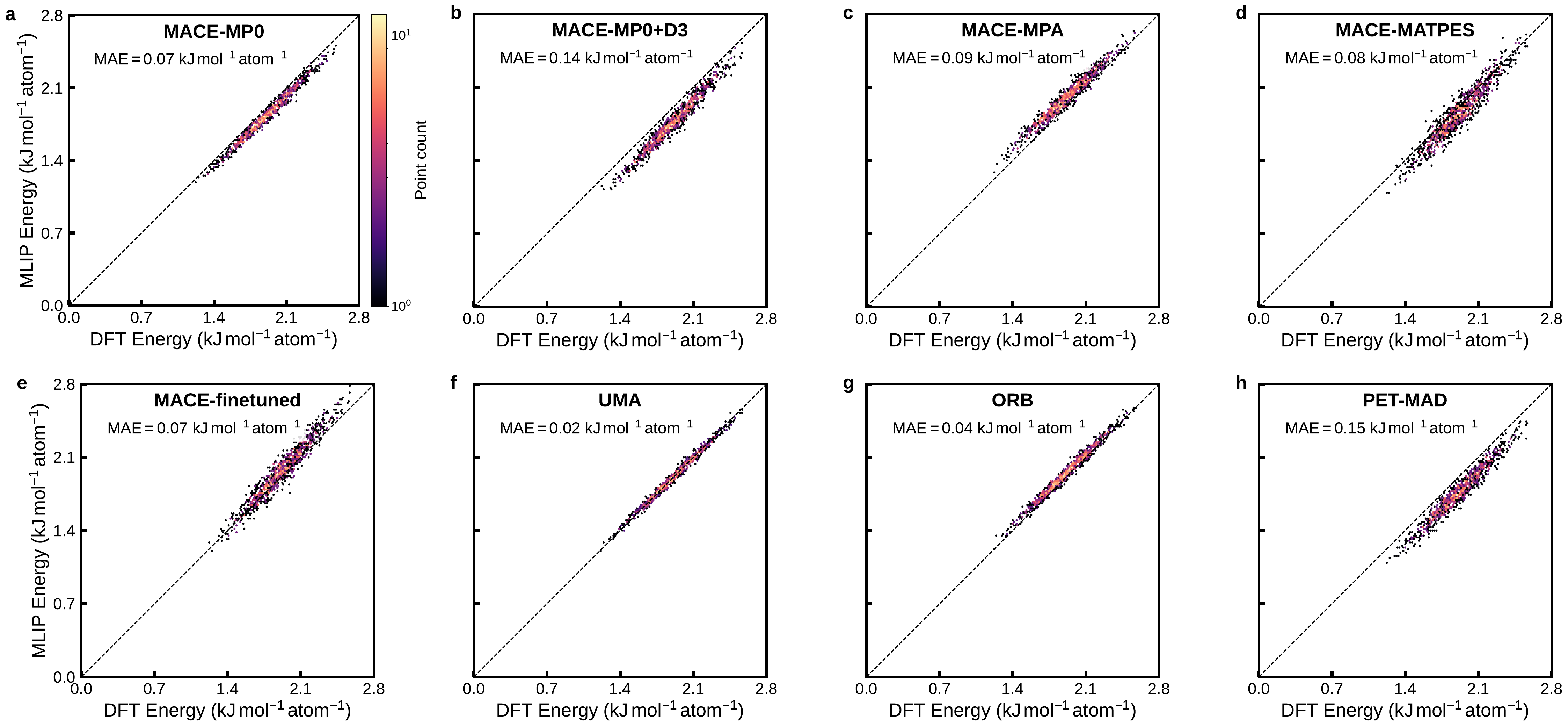}
    \caption{Parity plots for energies of the Li-zeolite-\ce{H2O} system, obtained with DFT (PBE+D3) and the suite of MLIPs used in this work. A total of 1100 structures were used for the parity plots, each with 601 atoms. All energies were referenced to the minimum energy structure calculated using DFT and the MLIPs.}
    \label{fig:energy_parity}
\end{figure}

\newpage
\section{Additional analysis on free energy reweighting}

\begin{figure}[H]
    \centering
    \includegraphics[width=0.5\linewidth]{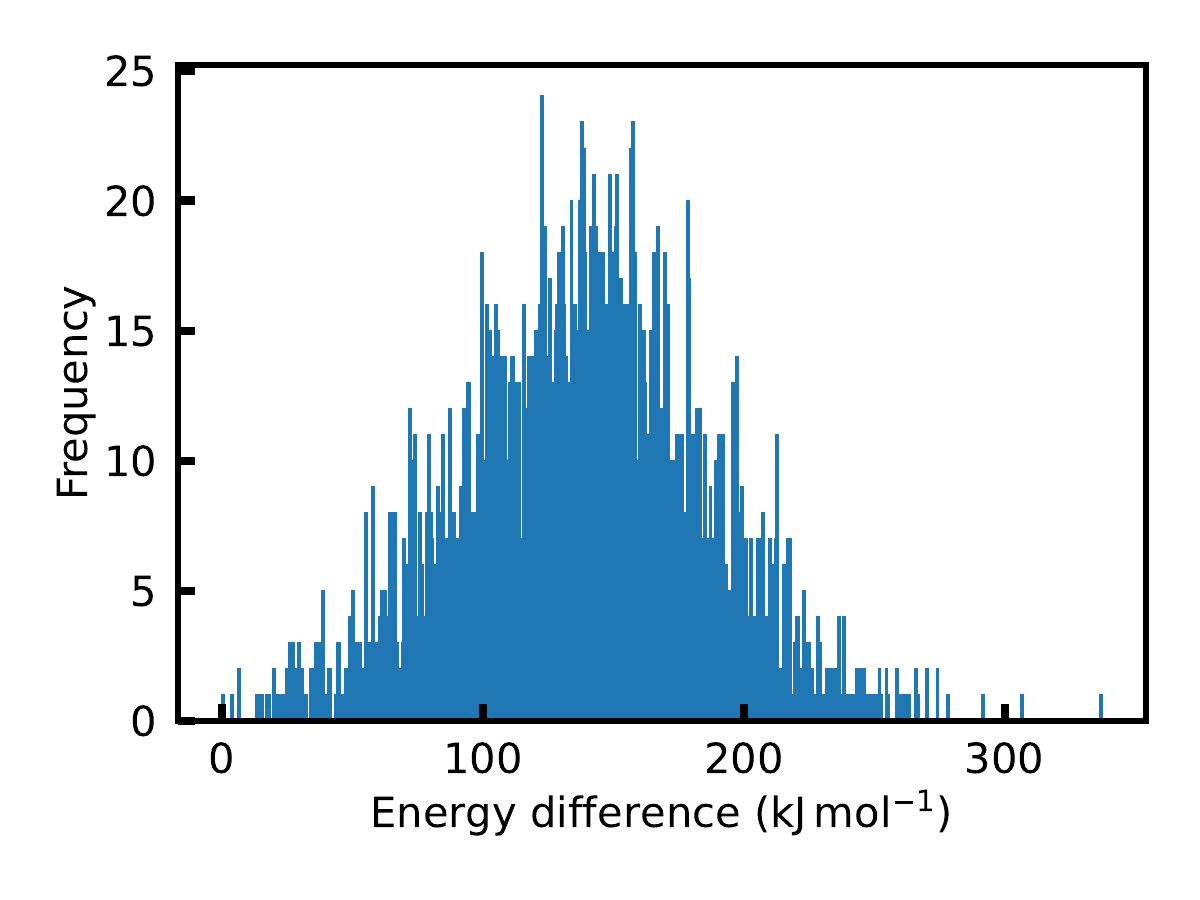}
    \includegraphics[width=0.5\linewidth]{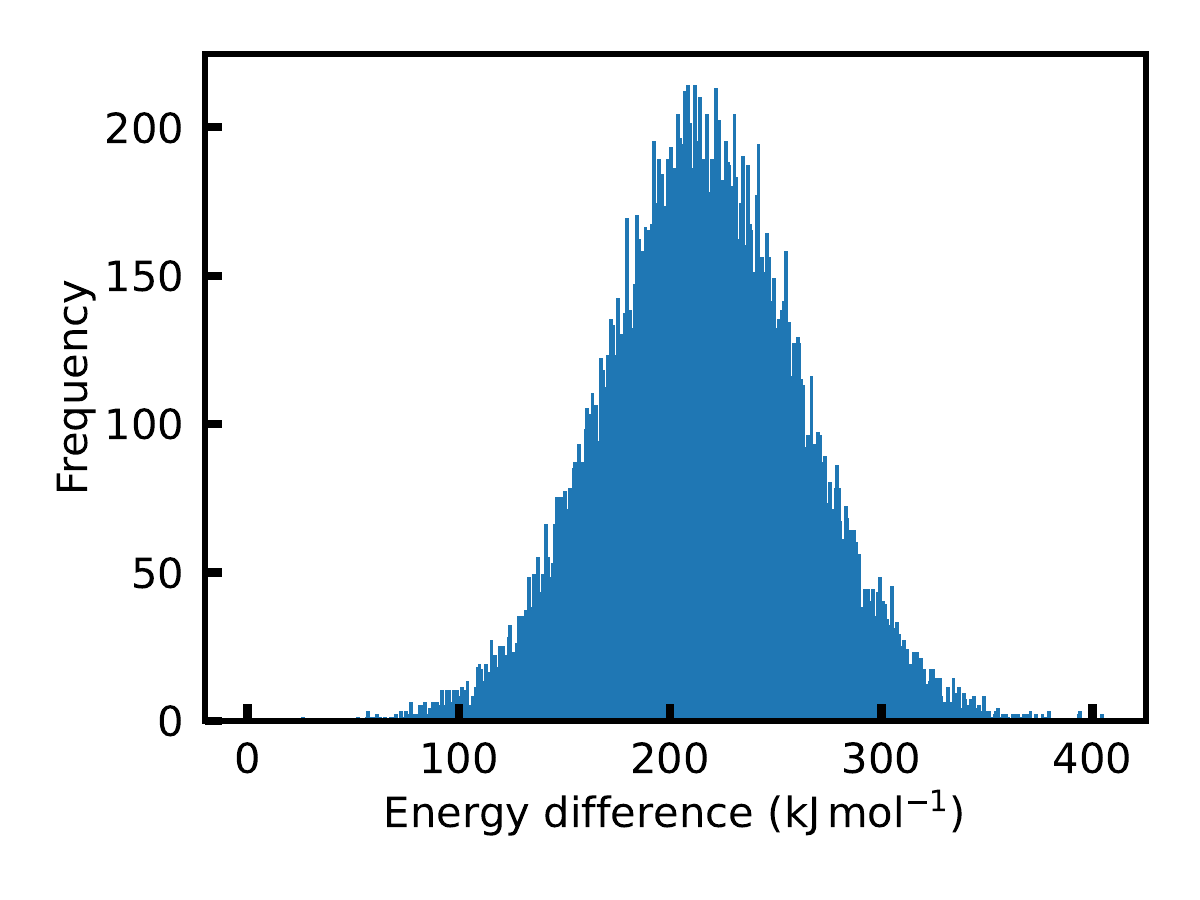}
    \includegraphics[width=0.5\linewidth]{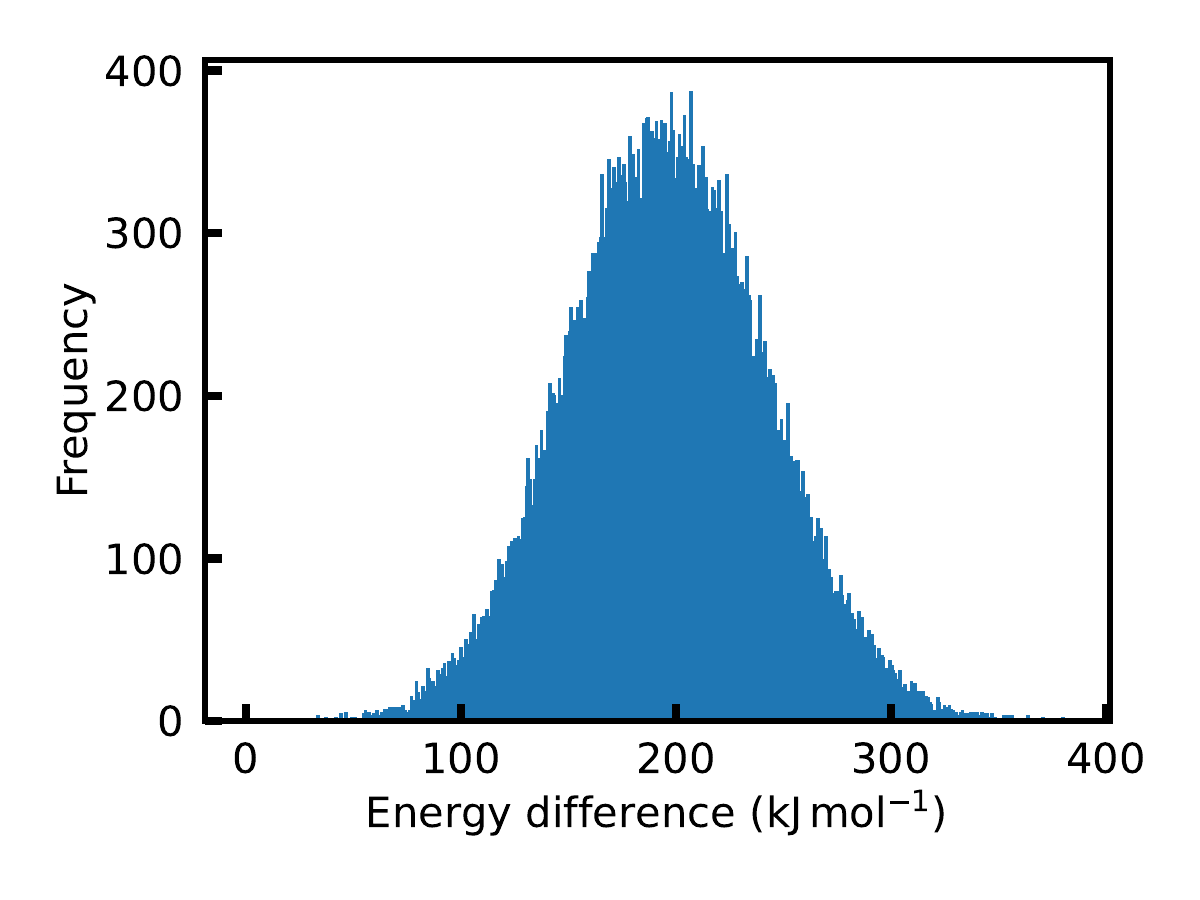}
    \caption{Frame-wise energy differences of the MACE-MATPES potential with the MACE-MP0 potential, with the smallest energy difference set to zero to eliminate an offset caused by the different reference energies of the two potentials. Each plot is given for a different number of random samples per window, going from 50, 500, to 1000 frames per umbrella window. The relative energy difference between the two potentials follows a Gaussian distribution, which gets smoother with the increase of frames per umbrella window. }
    \label{fig:placeholder}
\end{figure}

\begin{figure}[H]
    \centering
    \includegraphics[width=0.8\linewidth]{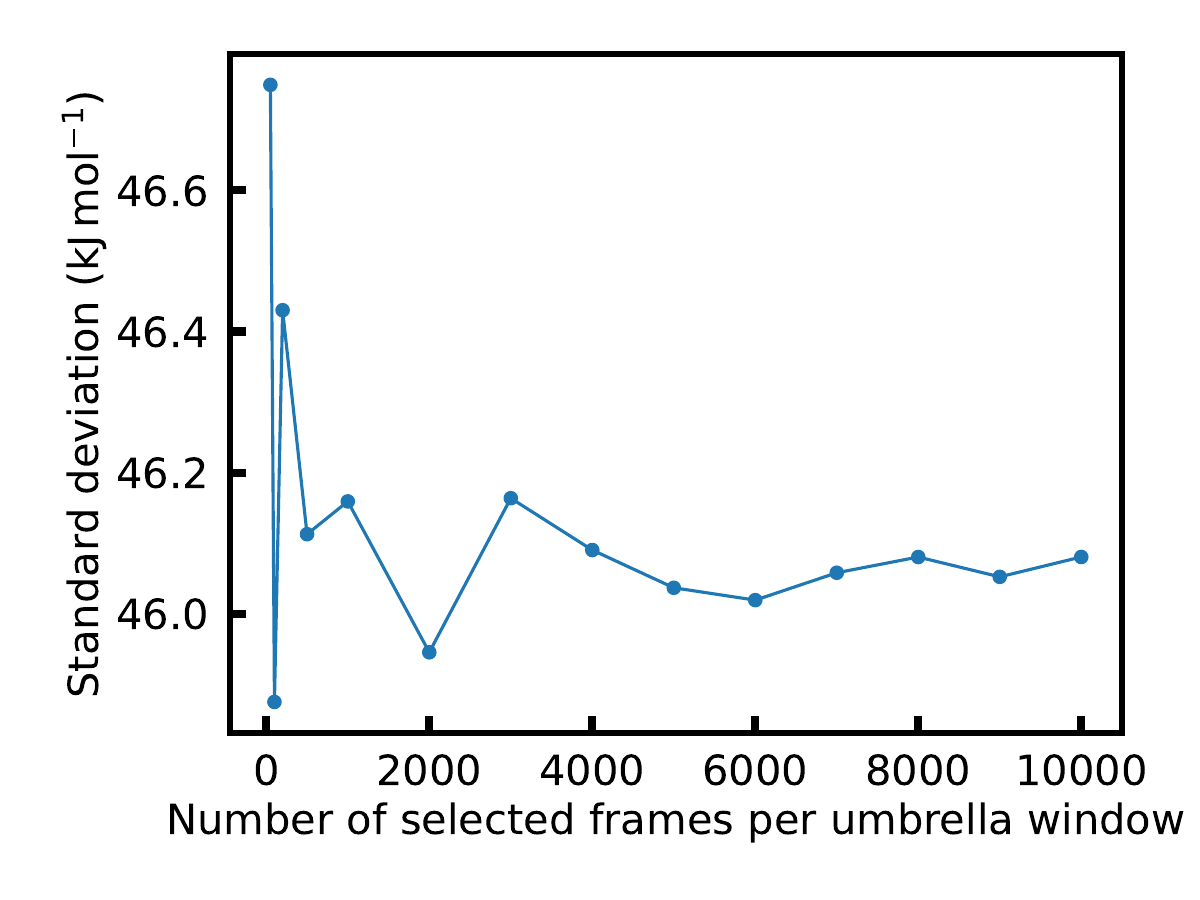}
    \caption{Standard deviations of the relative energy differences between MACE-MATPES and MACE-MP0 for different downsampling sizes per umbrella window. This illustrates that the standard deviation of the Gaussian distribution shown in Figure~S6 mostly remains consistent.}
    \label{fig:placeholder}
\end{figure}

\begin{figure}[H]
    \centering
    \includegraphics[width=0.8\linewidth]{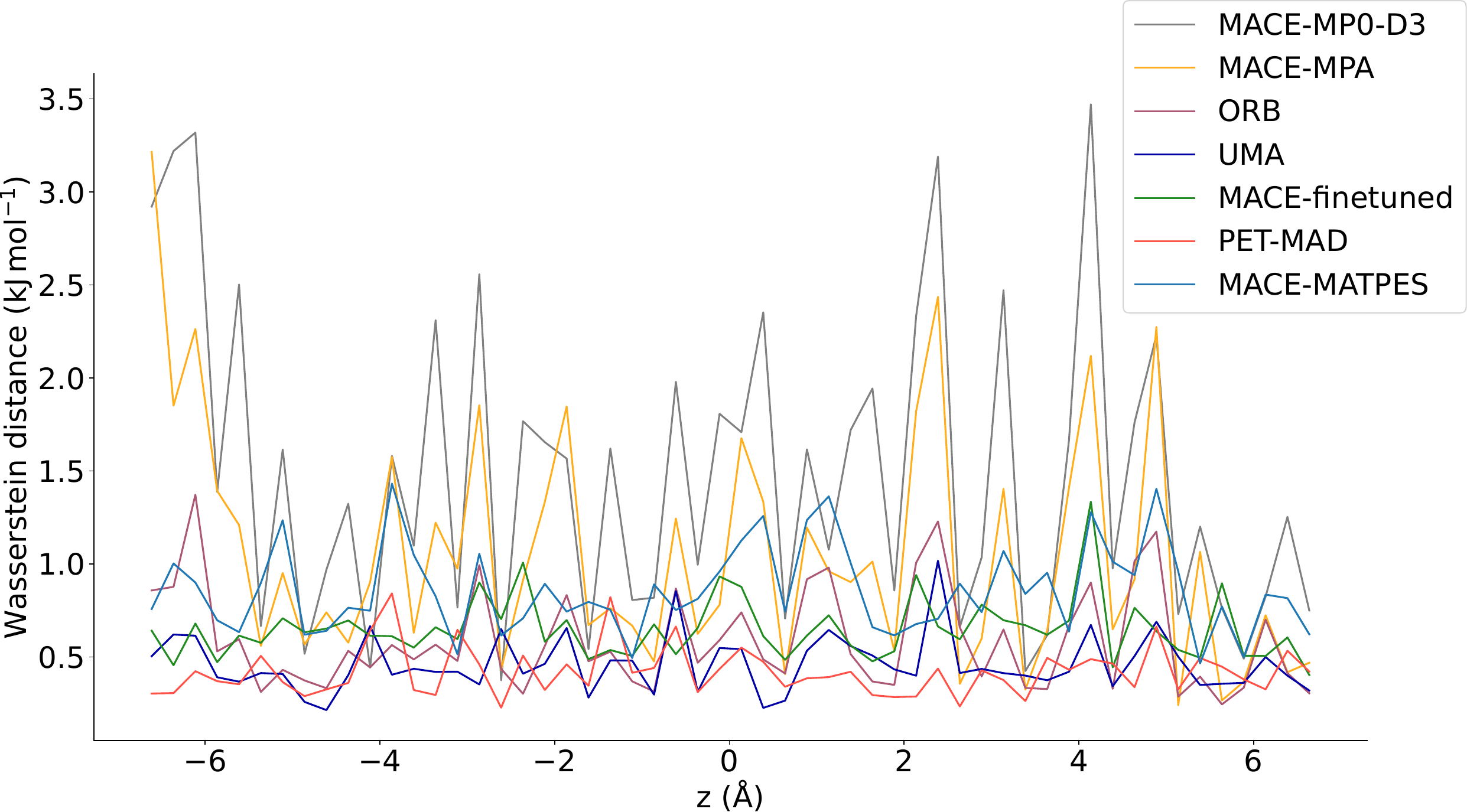}
    \caption{Wasserstein distance between the distribution of energy differences between the shown potential and MACE-MP0 in each umbrella window to a normal distribution with the same arithmetic mean and standard deviation. This is a measure of how much the relative energy differences deviate from a normal distribution.}
    \label{fig:placeholder}
\end{figure}

\begin{figure}[H]
    \centering
    \includegraphics[width=\linewidth]{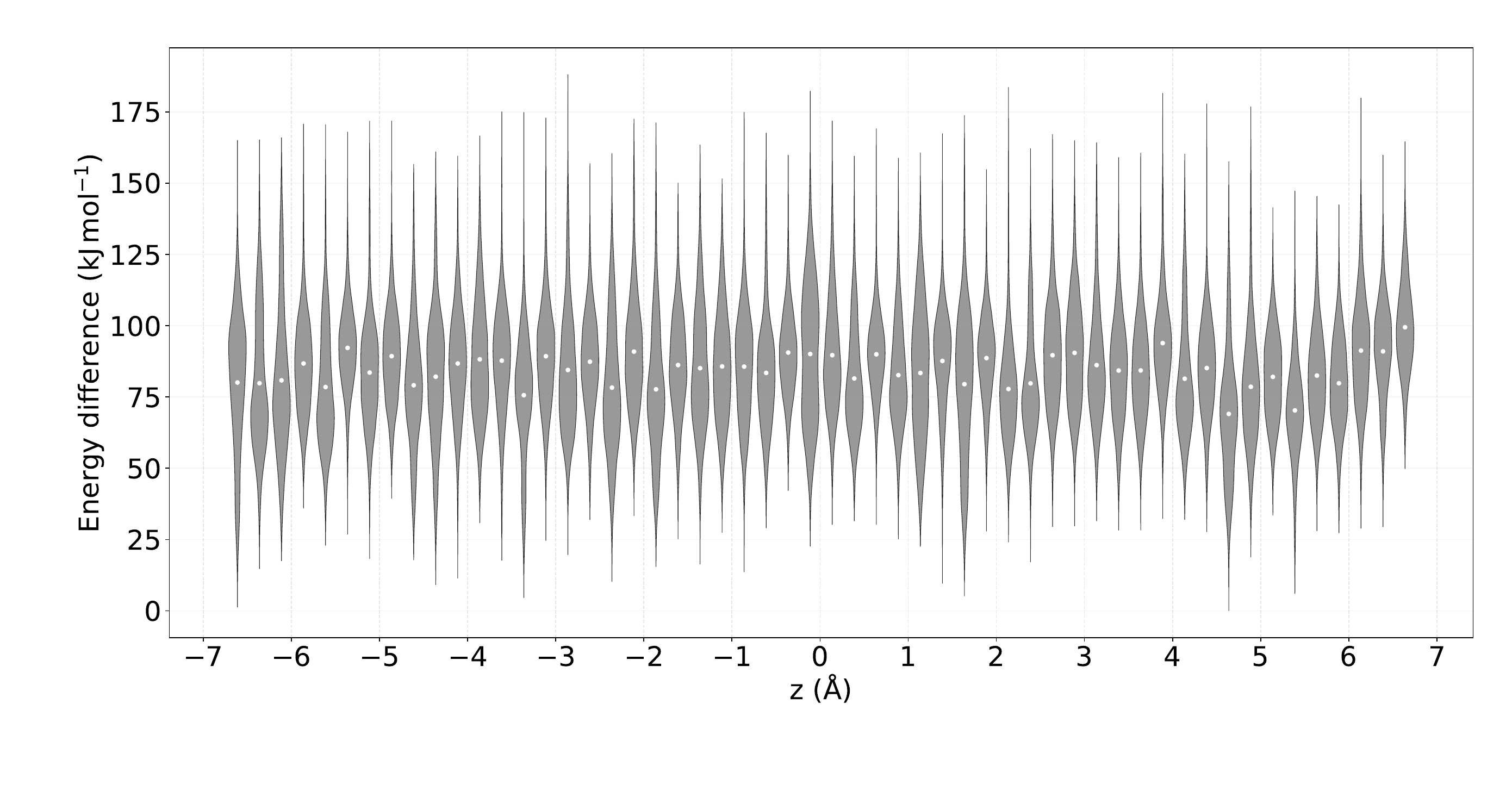}
    \includegraphics[width=\linewidth]{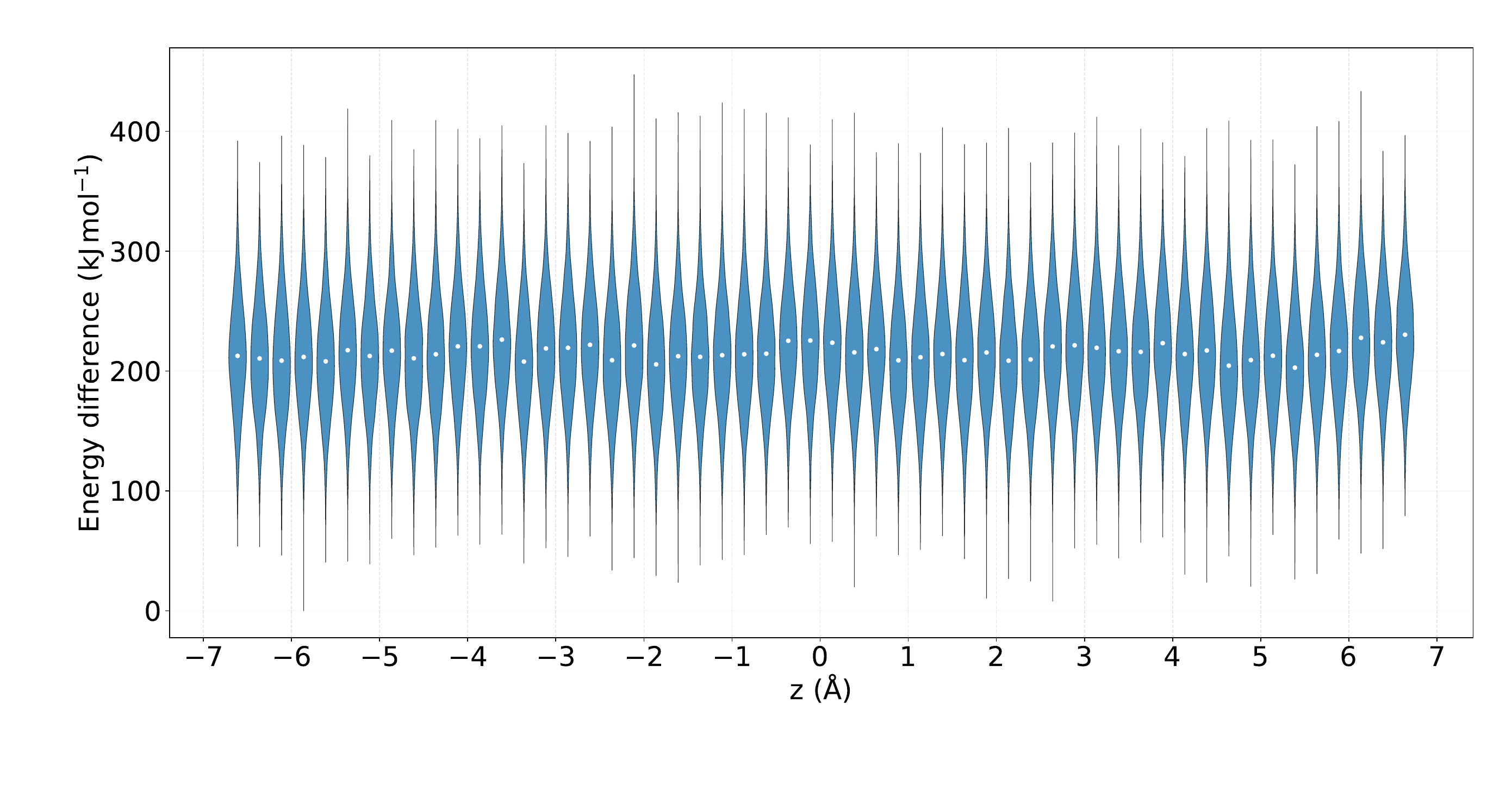}
    \caption{Frame-wise energy differences given for each separate umbrella window between MACE-MP0-D3 (top) and MACE-MATPES (bottom) with the MACE-MP0 potential. The smallest energy difference is set to zero to eliminate an offset caused by the different reference energies of the potentials.}
    \label{fig:placeholder}
\end{figure}

\section{Standard error estimates}
\subsection{Gaussian approximation for PMF reweighting}
The weighted mean and Bessel-corrected weighted variance of the dimensionless energy difference $\beta\Delta U$ within bin $z$ are estimated as
\begin{equation}\label{eq:weighted_mean}
    \mu_z = \sum_{i=1}^{N} p_z(i)\,\beta\Delta U(i) \ ,
\end{equation}
\begin{equation}\label{eq:weighted_var}
    \sigma_z^2 = \frac{\displaystyle\sum_{i=1}^{N} p_z(i)
                 \left(\beta\Delta U(i) - \mu_z\right)^2}
                {1 - \displaystyle\sum_{i=1}^{N} p_z^2(i)} \ ,
\end{equation}
where the denominator is the reliability correction for non-uniform weights. The standard errors on $\mu_z$ and  $\sigma_z^2$ are
\begin{equation}\label{eq:se_mu}
    \mathrm{SE}(\mu_z) = \sqrt{
        \sum_{i=1}^{N} p_z^2(i)
        \left(\beta\Delta U(i) - \mu_z\right)^2
    } \ ,
\end{equation}
\begin{equation}\label{eq:se_var}
    \mathrm{SE}(\sigma_z^2) =
    \frac{\sqrt{\displaystyle\sum_{i=1}^{N} p_z^2(i)
          \left[\left(\beta\Delta U(i) - \mu_z\right)^2
          - \sigma_z^2\right]^2}}
         {1 - \displaystyle\sum_{i=1}^{N} p_z^2(i)} \ .
\end{equation}
Propagating these through the reweighting estimate,
\begin{equation}\label{eq:gaussian_pmf}
\begin{split}
    A_\mathrm{B}(z) - A_\mathrm{A}(z) &= -k_B T \ln \langle e^{-\beta\Delta U}\rangle_{\mathrm{A},\xi=z} \\
    &= k_B T\, \frac{2\mu_{z}-\sigma_{z}^{2}}{2}.
\end{split}
\end{equation}

the uncertainty on the PMF correction at each $z$ is
\begin{equation}\label{eq:se_gaussian}
    \mathrm{SE}\!\left(\Delta A_z\right) =
    \frac{1}{\beta}
    \sqrt{
        \mathrm{SE}^2(\mu_z)
        +
        \frac{\mathrm{SE}^2(\sigma_z^2)}{4}
    } \ ,
\end{equation}
where the factor of $1/4$ arises from 
$\left(\partial\Delta A/\partial\sigma_z^2\right)^2 = (2\beta)^{-2}$.

\newpage
\subsection{Energy-only approximation}
To quantify the statistical uncertainty on the energy-only estimate of PMF, we note that the correction in this estimate,

\begin{equation}\label{eq:energy_only}
    A_\mathrm{B}(z) \approx A_\mathrm{A}(z) + \langle \Delta U \rangle_{\mathrm{A},\xi=z}.
\end{equation}

is simply a weighted mean, so its standard error follows directly from error propagation of the weighted mean with MBAR:
\begin{equation}\label{eq:se_energyonly}
    \mathrm{SE}\!\left(\langle\Delta U\rangle_{\mathrm{A},\xi=z}\right) =
    \sqrt{\sum_{i=1}^{N} p_z^2(i)
    \left(\Delta U(i) - \langle\Delta U\rangle_{\mathrm{A},\xi=z}\right)^2} \ .
\end{equation}
This standard error reflects only the \emph{statistical} precision of estimating the mean energy gap from a finite ensemble. It does not capture the systematic bias inherent to the energy-only approximation 
itself: from eqn~\eqref{eq:gaussian_pmf}, the neglected entropic contribution is $k_BT\sigma_z^2/2$ per bin, which grows with the width of the $\Delta U$ distribution. The $\pm\mathrm{SE}$ band therefore represents a lower bound on the true uncertainty and should not be interpreted as capturing the full error of the approximation.

\begin{figure}
    \centering
    \includegraphics[width=\linewidth]{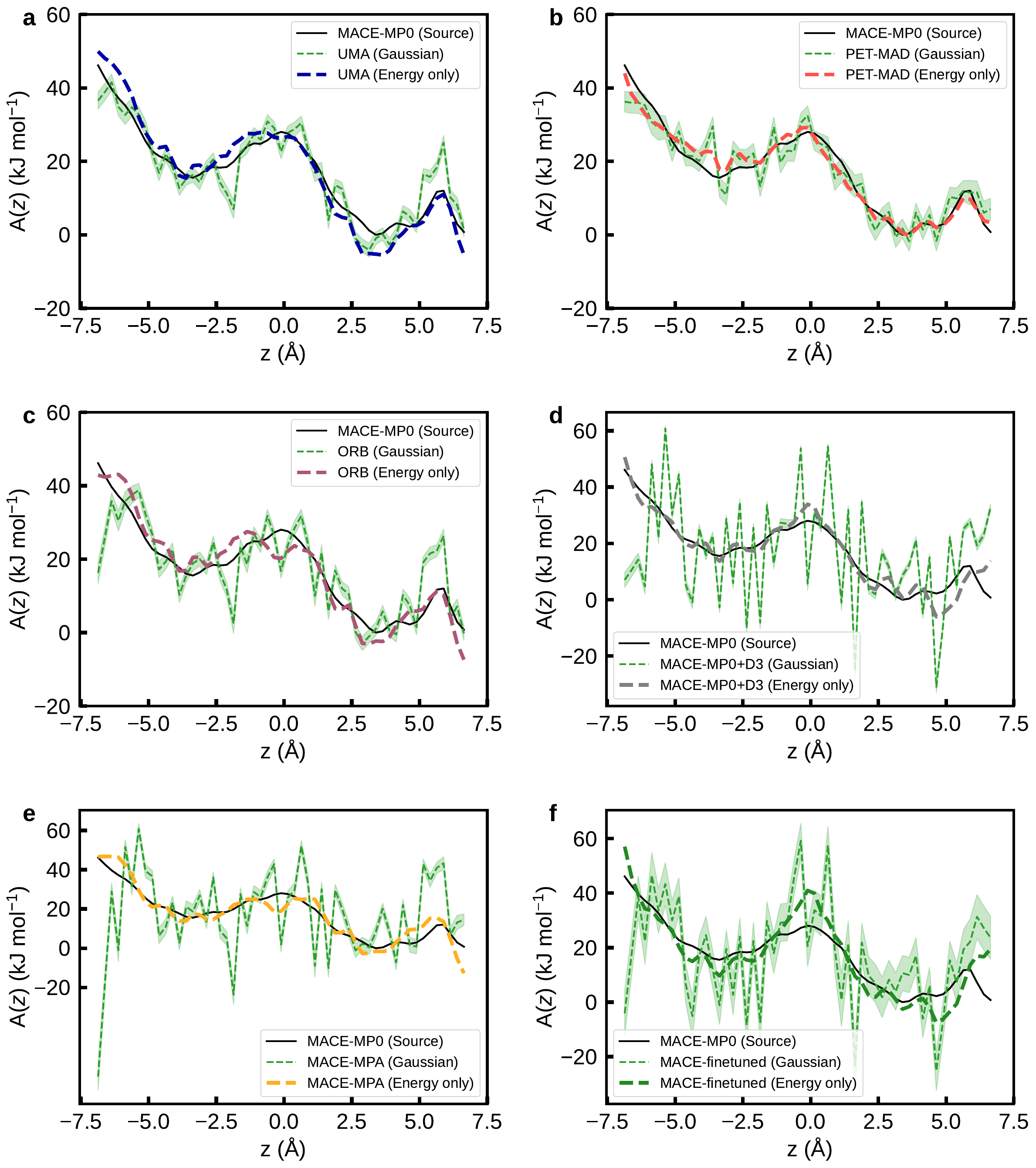}
    \caption{Reweighted PMFs obtained using the Gaussian and the energy-only approximation ($\langle \Delta U \rangle_{A,\xi=z}$) for six MLIPs, (a) UMA, (b) PET-MAD, (c) ORB, (d) MACE-MP0-D3, (e) MACE-MPA, and (f) MACE-finetuned, starting from the MACE-MP0 umbrella sampling trajectory (black). Dashed lines show the reweighted PMF for each target MLIP; the shaded areas indicate twice the standard error, computed with eqn~\eqref{eq:se_gaussian}.}
    \label{fig:placeholder}
\end{figure}

\section{Reweighting and its limits}

Suppose one has performed biased or unbiased simulations on a potential $U^\mathrm{A}$ and wishes to obtain the PMF for a different PES $U^\mathrm{B}$ without running new, expensive simulations. 
One approach is to \emph{reweight} the frames already sampled under $U^\mathrm{A}$:
\begin{equation}
    W_\mathrm{rw}(\mathbf{x}) 
    = e^{-\beta \Delta U(\mathbf{x})}\, W(\mathbf{x}) \ ,
\end{equation}
where $\Delta U = U^\mathrm{B} - U^\mathrm{A}$ and $W(\mathbf{x})$ are the weights for each configuration obtained from simulations on $U^\mathrm{A}$. 
For unbiased simulations, all weights $W$ are equal; for umbrella sampling, the unbiased weights $W$ are obtained with MBAR (see main text). 
The reweighted PMF for $U^\mathrm{B}$ is then
\begin{align}
    A^\mathrm{B}(z) 
    & = - k_\mathrm{B}T \ln \rho^\mathrm{B}(z) 
      = - k_\mathrm{B}T \ln 
        \bigl\langle \delta(\xi(\mathbf{x}) - z)\bigr\rangle_\mathrm{B} 
      \nonumber\\
    & = - k_\mathrm{B}T \ln 
      \sum_k^K \sum_j^{N_k} W_\mathrm{rw}(k,j)\, 
      \delta\bigl(\xi(k,j) - z\bigr) \ ,
\end{align}
where $\langle \cdot \rangle_\mathrm{B}$ denotes an ensemble average over samples drawn from the ensemble associated with $U^\mathrm{B}$, the tuple $(k,j)$ denotes frame $j$ from simulation window $k$, and $\xi(\cdot)$ is the collective variable.

The straightforward reweighting formula, however, conceals the fact that the individual weights $W_\mathrm{rw}$ can vary by many orders of magnitude due to the exponential factor $e^{-\beta\Delta U}$. 
It is therefore more instructive to express the new PMF as a correction to the original one. 
Using the identity $\langle O \rangle_\mathrm{B} \propto \langle O\,e^{-\beta\Delta U}\rangle_\mathrm{A}$ and factoring out the conditional average at fixed $\xi = z$:
\begin{align}
    A^\mathrm{B}(z) 
    & = - k_\mathrm{B}T \ln \Bigl(
        \bigl\langle \delta(\xi(\mathbf{x}) - z)
        \bigr\rangle_\mathrm{A}\,
        \bigl\langle e^{-\beta \Delta U}
        \bigr\rangle_{\mathrm{A},\,\xi=z}
      \Bigr) + \mathrm{const} \ , \\[4pt]
    \Delta A(z) = A^\mathrm{B}(z) - A^\mathrm{A}(z) 
    & = - k_\mathrm{B}T \ln 
      \bigl\langle e^{-\beta \Delta U}\bigr\rangle_{\mathrm{A},\,\xi=z} 
      + \mathrm{const} \ .
    \label{eq:pmf_correction}
\end{align}
The subscript on each ensemble average indicates the PES from which samples were drawn; $\xi = z$ further restricts the average to frames falling in the bin centered on $z$. 
The additive constant — the global free energy difference $k_\mathrm{B}T\ln\langle e^{-\beta\Delta U}\rangle_\mathrm{A}$ — is independent of $z$ and does not affect the shape of the PMF; it is absorbed into the overall normalization.

Unfortunately, the conditional average $\langle e^{-\beta \Delta U}\rangle_{\mathrm{A},\,\xi=z}$ is notoriously ill-conditioned. 
To see why, we recast it as an integral over the distribution of energy differences:
\begin{equation}
    \bigl\langle e^{-\beta \Delta U}\bigr\rangle_\mathrm{A} 
    = \int_{-\infty}^\infty \mathrm{d}(\beta\Delta U)\;
      e^{-\beta \Delta U}\, P_\mathrm{A}(\beta\Delta U) \ ,
    \label{eq:integral_deltaU}
\end{equation}
where $P_\mathrm{A}(\beta\Delta U)$ is the distribution of dimensionless energy differences sampled under $U^\mathrm{A}$ (the bin index $z$ is suppressed here because the difficulty is present in every bin). 
If the system has sufficiently many degrees of freedom, the central limit theorem justifies approximating $P_\mathrm{A}$ as a Gaussian:
\begin{equation}
    P_\mathrm{A}(\beta\Delta U) \approx 
    \mathcal{N}(\beta\Delta U;\, \mu,\, \sigma^2) \ .
\end{equation}

Evaluating Eq.~(\ref{eq:integral_deltaU}) numerically is equivalent to estimating the integral by Monte Carlo, drawing samples from $P_\mathrm{A}(\beta\Delta U)$ and weighting each by $e^{-\beta\Delta U}$. 
The difficulty is that this exponential factor grows as $\beta\Delta U$ decreases, so the dominant contribution to the integral comes from the left tail of $P_\mathrm{A}$ -- precisely the region where samples are exponentially rare. 
The bulk of the available frames, which cluster near the mean $\mu$, contribute comparatively little.

To make this quantitative, we note that roughly 99\% of all samples fall within $3\sigma$ of $\mu$, and compute the fraction of the true integral captured by this region:
\begin{equation}
    \frac{%
        \displaystyle\int_{\mu-3\sigma}^{\mu+3\sigma} 
        \mathrm{d}(\beta\Delta U)\;
        e^{-\beta \Delta U}\, P_\mathrm{A}(\beta\Delta U)}{%
        \displaystyle\int_{-\infty}^\infty 
        \mathrm{d}(\beta\Delta U)\;
        e^{-\beta \Delta U}\, P_\mathrm{A}(\beta\Delta U)}
    = \frac{1}{2} \left(
        \mathrm{erf}\!\left(\frac{\sigma + 3}{\sqrt{2}}\right) 
        - \mathrm{erf}\!\left(\frac{\sigma - 3}{\sqrt{2}}\right)
      \right) \ ,
    \label{eq:coverage}
\end{equation}
where $\sigma$ dimensionless or expressed in multiples of $\beta^{-1} = k_\mathrm{B}T$.
The covered fraction is close to unity for $\sigma \le 1$, falls to roughly 0.5 at $\sigma \approx 3$, and essentially vanishes for $\sigma > 5$. 
A broad $\Delta U$ distribution therefore leads to a severely undersampled integral and an unreliable free energy estimate, regardless of the total number of simulation frames available.

\section{Scaling with System Size}

To illustrate how the difficulty of reweighting grows with the number of degrees of freedom, we consider a system in which every degree of freedom is governed by a harmonic potential with the same force constant $k$, and in which the two PES differ only in the equilibrium position of every mode by a uniform shift $x_0$. 
Concretely, we take
\begin{align}
    U^\mathrm{A}(\mathbf{x}) 
    & = \frac{N}{\beta}\ln \sqrt{\frac{2\pi}{k\beta}} 
      + \frac{k}{2}\, \mathbf{x}^\top \mathbf{x} 
      = \frac{N}{\beta}\ln \sqrt{\frac{2\pi}{k\beta}} 
      + \frac{k}{2} \sum_{i=1}^N x_i^2 \ , \\[4pt]
    U^\mathrm{B}(\mathbf{x}) 
    & = \frac{N}{\beta}\ln \sqrt{\frac{2\pi}{k\beta}} 
      + \frac{k}{2}\,(\mathbf{x} - \mathbf{x}_0)^\top(\mathbf{x} - \mathbf{x}_0)
      = \frac{N}{\beta}\ln \sqrt{\frac{2\pi}{k\beta}} 
      + \frac{k}{2} \sum_{i=1}^N (x_i - x_0)^2 \ ,
\end{align}
where $N$ is the number of degrees of freedom and $\mathbf{x}_0 = x_0\,\mathbf{1}$ is the uniform shift vector. 
Internal coordinates are chosen so that the minimum of $U^\mathrm{A}$ coincides with $x_i = 0$ for all $i$. 
The prefactor $\frac{N}{\beta}\ln\sqrt{\frac{2\pi}{k\beta}}$ is a normalization constant ensuring that $e^{-\beta U}$ integrates to unity.

For a configuration $\mathbf{x}$, the potential energy difference is
\begin{align}
    \Delta U(\mathbf{x}) 
    & = U^\mathrm{B}(\mathbf{x}) - U^\mathrm{A}(\mathbf{x}) \nonumber \\
    & = kx_0 \sum_{i=1}^N \left(\frac{x_0}{2} - x_i\right) \ .
\end{align}
The mean energy difference is
\begin{align}
    \langle \Delta U\rangle_\mathrm{A} 
    & = \left(\frac{2\pi}{k\beta}\right)^{-N/2} kx_0 
      \int \mathrm{d}\mathbf{x}\; 
      e^{-\frac{\beta k}{2}\mathbf{x}^\top\mathbf{x}}\; 
      \sum_{i=1}^N \left(\frac{x_0}{2} - x_i\right) \nonumber \\
    & = \frac{Nkx_0^2}{2} \ ,
\end{align}
where we used $\langle x_i \rangle_\mathrm{A} = 0$.
The second moment is
\begin{align}
    \langle (\Delta U)^2\rangle_\mathrm{A} 
    & = \left(\frac{2\pi}{k\beta}\right)^{-N/2} k^2x_0^2 
      \int \mathrm{d}\mathbf{x}\; 
      e^{-\frac{\beta k}{2}\mathbf{x}^\top \mathbf{x}}\; 
      \sum_{i=1}^N\sum_{j=1}^N 
      \left(\frac{x_0}{2} - x_i\right)\!\left(\frac{x_0}{2} - x_j\right) 
      \nonumber \\
    & = k^2x_0^2 \left( \frac{N^2 x_0^2}{4} + \frac{N}{\beta k}\right) \ ,
\end{align}
where we used $\langle x_ix_j\rangle_\mathrm{A} = \delta_{ij}/(\beta k)$ (independent harmonic modes).
The mean and variance of $P_\mathrm{A}(\beta\Delta U) = \mathcal{N}(\beta\Delta U;\,\mu,\,\sigma^2)$ are therefore
\begin{align}
    \mu    & = \beta\langle \Delta U\rangle_\mathrm{A} 
             = \frac{N\beta kx_0^2}{2} \ , \\[4pt]
    \sigma^2 
           & = \beta^2\!\left(\langle(\Delta U)^2\rangle_\mathrm{A} 
             - \langle\Delta U\rangle_\mathrm{A}^2\right) 
             = \beta k x_0^2 N \ .
\end{align}
The key observation is that $\sigma = x_0\sqrt{\beta k N}$ grows as $\sqrt{N}$. 
As established in the previous section, reliable reweighting requires $\sigma \le 3$. 
No matter how weak the force constant $k$, how small the shift $x_0$, or how high the temperature (small $\beta$), increasing $N$ will eventually push $\sigma$ beyond this threshold. 
Reweighting between two PES is therefore an intrinsically size-limited procedure.

\subsection{Back-of-the-Envelope Estimate}

A purely harmonic model is a reasonable first approximation for stiff covalent solids such as silica or diamond. 
A representative C--C bond force constant in diamond is $k = 5.5\times10^{-18}$~J\,\AA$^{-2}$. 
At $T = 450$~K (the temperature used in the simulations of this work) $\beta \approx 1.6\times10^{20}$~J$^{-1}$, giving $\sqrt{\beta k} \approx 30$~\AA$^{-1}$.

Suppose the two force fields differ by only $x_0 = 1$~pm $= 0.01$~\AA\ in equilibrium bond length — a very small discrepancy. 
Then $\sigma \approx 0.3\sqrt{N}$, and the condition $\sigma \le 3$ limits reliable reweighting to systems with fewer than $N \approx 100$ internal degrees of freedom. 
For a more substantial disagreement of $x_0 = 0.1$~\AA, even a two-degree-of-freedom system would be problematic: $\sigma = 3\sqrt{2} \approx 4.24$, already placing the reweighting in the unreliable regime identified by eqn.~\eqref{eq:coverage}.

\bibliography{sn-bibliography}